\documentclass[reqno]{amsart}
\usepackage{amsmath,amssymb,graphicx}
\newtheorem{theorem}{Theorem}[section]   
\newtheorem{proposition}{Proposition}[section]   
\newtheorem{lemma}{Lemma}[section]   
\newtheorem{corollary}{Corollary}[section]        

\newtheorem{remark}{Remark}[section]   

\numberwithin{equation}{section}
\numberwithin{theorem}{section}
\numberwithin{proposition}{section}
\numberwithin{lemma}{section}
\numberwithin{corollary}{section}
\numberwithin{remark}{section}

\newcommand{\mc}[1]{{\mathcal #1}}
\newcommand{\bb}[1]{{\mathbb #1}}
\newcommand{\rme}{\mathrm{e}}

\newcommand{\rmd}{\mathrm{d}}

\let\a=\alpha \let\b=\beta  \let\g=\gamma  \let\d=\delta \let\e=\varepsilon
\let\z=\zeta      \let\k=\kappa \let\l=\lambda
\let\m=\mu        \let\x=\xi     \let\p=\pi    \let\r=\rho
\let\s=\sigma \let\t=\tau    \let\c=\chi
   \let\o=\omega
\let\G=\Gamma   \let\L=\Lambda

 \def\BBB{{\mathcal B}}

\def\be{\begin{equation}}
\def\ee{\end{equation}}
\def\bp{\begin{pmatrix}}
\def\ep{\end{pmatrix}}
\def\bea{\begin{eqnarray}}
\def\eea{\end{eqnarray}}

\def\\{\par\medskip}
\let\io=\infty
\let\0=\noindent
\def\media#1{{\langle#1\rangle}}
\let\dpr=\partial
\def\sign{{\rm sign}}
\def\const{({\rm const.})}
\def\lis{\overline}

\def\dist{{\rm{dist}}}

\begin{document}

\title[Froth-like minimizers of a non local free energy functional]{Froth-like minimizers of a non local free energy functional with competing interactions}

\author [P.\ Butt\`a] {Paolo Butt\`a}
\address{Paolo Butt\`a, Dipartimento di Matematica, SAPIENZA Universit\`a di Roma, P.le Aldo Moro 5, 00185 Roma, Italy}
\email{butta@mat.uniroma1.it}

\author [R.\ Esposito] {Raffaele Esposito}
\address{Raffaele Esposito, M\&MOCS, Universit\`a dell'Aquila, Cisterna di Latina, 04012 Italy}
\email{raff.esposito@gmail.com}

\author [A.\ Giuliani] {Alessandro Giuliani}
\address{Alessandro Giuliani, Dipartimento di Matematica, Universit\`a di Roma Tre, Largo S. Leonardo Murialdo 1, 00146 Roma, Italy} 
\email{giuliani@mat.uniroma3.it} 

\author [R.\ Marra] {Rossana Marra}
\address{Rossana Marra, Dipartimento di Fisica and Unit\`a INFN, Universit\`a di Roma Tor Vergata, Via della Ricerca Scientifica 1, 00133 Roma, Italy}
\email{marra@roma2.infn.it}

\begin{abstract}
We investigate the ground and low energy states of a one dimensional non local free energy functional describing at a mean field level a spin system with both ferromagnetic and antiferromagnetic interactions. In particular, the antiferromagnetic interaction is assumed to have a range much larger than the ferromagnetic one. The competition between these two effects is expected to lead to the spontaneous emergence of a regular alternation of long intervals on which the spin profile is magnetized either up or down, with an oscillation scale intermediate between the range of the ferromagnetic and that of the antiferromagnetic interaction. In this sense, the optimal or quasi-optimal profiles are ``froth-like": if seen on the scale of the antiferromagnetic potential they look neutral, but if seen at the microscope they actually consist of big bubbles of two different phases alternating among each other. In this paper we prove the validity of this picture, we compute the oscillation scale of the quasi-optimal profiles and we quantify their distance in norm from a reference periodic profile. The proof consists of two main steps: we first coarse grain the system on a scale intermediate between the range of the ferromagnetic potential and the expected optimal oscillation scale; in this way we reduce the original functional to an effective ``sharp interface" one. Next, we study the latter by reflection positivity methods, which require as a key ingredient the exact locality of the short range term. Our proof has the conceptual interest of combining coarse graining with reflection positivity methods, an idea that is presumably useful in much more general contexts than the one studied here.
\end{abstract}

\maketitle
\thispagestyle{empty}

%%%%%%%%%%%%%%%%%%%%%%%%%%%%%%%%%%%%%%%%%%%%%%%%%%%
\section{Introduction and description of the model}
\label{sec:1}
%%%%%%%%%%%%%%%%%%%%%%%%%%%%%%%%%%%%%%%%%%%%%%%%%%%

The competition between short-range attractive forces and long-range dipolar forces can give rise to the spontaneous formation of periodic patterns, such as stripes or bubbles, as observed in several quasi two-dimensional (2D) systems, e.g., micromagnets and magnetic films, ferrofluids, quasi 2D electron gases and high temperature superconductors, liquid crystals, system of suspended lipidic molecules on the water surface, assemblies of diblock copolymers, martensitic phase transitions; see, e.g., \cite{ref5,ref4, ref6, ref7, ref8, ref9, ref10, ref11, ref12, ref13, ref14, ref15}. From a mathematical point of view, these systems are modelled by a microscopic or mesoscopic non-convex energy functional, whose low energy states are expected to display the same pattern formation phenomenon. There are a number of rigorous indications of the emergence of regular structures, ranging from equipartition to rigorous upper and lower bounds on the minimizing energy \cite{ACO,BCDM,BL,C,CCKO,C1,C2,CO,DKOM,GLL11,KM1,KM2}.  In a few cases, the existence of periodic ground states can be rigorously proved \cite{ref16,ref17,ref18,GLL1,GLL07,GLL2,GLL3,GMu,M,ref20,RW,ref21}. Among these, a one dimensional (1D) Ising model with nearest neighbor ferromagnetic (FM) exchange and long range power law antiferromagnetic (AF) interaction, where periodicity of the ground states was proved by means of a generalized reflection positivity (RP) method \cite{GLL1}. Later, such proof of periodicity was extended to other systems, both in one and two dimensions, in the discrete or continuum setting \cite{GLL07,GLL2,GLL3,GLL11,GMu}; in particular, we mention two continuous versions of the 1D spin model studied in \cite{GLL1}, where the discrete spin Hamiltonian is replaced by an effective free energy functional and the configuration of discrete Ising spins is  replaced by a magnetization profile $\s(x)$ with $x\in\mathbb R$, either assuming all possible values between $-1$ and $1$ (the ``soft spin" case, see \cite{GLL2}), or assuming only values $\pm1$ (the ``Ising spin" case, see \cite{GLL3}). In both cases, a crucial technical assumption for the method of the proof to work is that the short range FM term appearing in the free energy functional is exactly local, i.e., it is modeled by a gradient term or by a local surface tension term, depending on whether one considers the soft or Ising spin case. Under this assumption, the minimizers are exactly periodic and consist of intervals of constant length $h^*$ (the optimal modulation length) in which the magnetization has constant sign, the sign oscillating from plus to minus or viceversa when one moves from a given interval to the following one. Moreover, the magnetization profiles with free energy close to the minimal one are very close to the periodic minimizers.

From a physical point of view, the locality of the surface tension term is a phenomenological (often unjustified) assumption and it should be essentially irrelevant as far as the results are concerned. In other words, if we replace a local surface tension term by a short but finite ranged one, with range much smaller than the range of the AF interaction, the magnetization profiles minimizing the free energy functional, or with free energy sufficiently close to the minimum, should still consist of a regular alternation of intervals where the magnetization is positive or negative. The exact periodicity of the minimizers may be a special feature of models with a local surface tension, but approximate periodicity should be a robust property. Therefore, it is important to understand whether the results of \cite{GLL2,GLL3} can be extended to cases where the generalized RP method breaks down, due to the non locality of the short range interaction. The extension has, on the one hand, a specific interest for the class of 1D magnetic models we are considering: in fact, we are not aware of examples of free energy functionals with strictly local penalization of gradients which can be directly derived as continuum limit of microscopic particle models. On the other hand, it has a more general conceptual importance: it is of great interest to develop methods allowing one to extend the validity of results based on RP to cases where RP does not hold exactly, possibly by combining it with coarse graining or averaging methods.
 
In this paper we attempt a first extension in this direction, by focusing on a free energy functional that arises naturally in 1D Ising models with competing long range interactions at positive temperature in a specific mean field limit, known as {\it Kac limit}. To be more precise, let us consider the 1D spin system described by the Hamiltonian 
\be 
\label{00.1}
H^{(\e)}_{\e^{-1}L}=-\e\sum_{x<y}J(\e(x-y)) \, \s_x\s_y +\g\e\sum_{x<y}v(\g\e(x-y))\,\s_x\s_y \;,
\ee
where $\g$ and $\e$ should be thought of as small parameters, the sums over $x$ and $y$ run over the set $\mathbb Z\cap [0,\e^{-1}L]$, $J(x)$ is an even, nonnegative, smooth monotone function with support equal to $[-1,1]$ and $v(x)$ is smooth and reflection positive, that is: $v$ can be written as the Laplace transform of a positive measure, $v(x)=\l\int_0^\io\!\m(\rmd\a) \,\rme^{-\a |x|}$, with $ \m(\rmd\a)$ a probability measure on $[0,\io)$ such that $\l\int_0^\io\!\m(\rmd\a)\,\a^{-1}=\int_0^\io\!\rmd x\, v(x)<\io$, $\l$ a positive constant. We remark that this implies $v$ is a $C^\io$ function on $\mathbb R\setminus \{0\}$, with derivatives of all orders extendable up to $x=0$. To simplify some technical points in the following, we shall further assume that $\m(d\a)$ has compact support, well separated from $0$ (that is, $\g v(\g x)$ is a superposition of exponentials, with range comparable with $\g^{-1}$).
Given $\b>0$ and $\e\ll1$, on a coarse grained scale of the order $\e^{-1}$, the typical configurations with respect to the Gibbs measure with Hamiltonian Eq.\eqref{00.1} and inverse temperature $\b$ are described by a nonlocal large deviation functional $\mc E^{(\g)}_{[0,L]}[\phi]$. Roughly speaking, this means that the probability of the spin configurations compatible with the coarse grained profile $\phi$ is approximately given  by $\exp\{-\b\e^{-1}\mc E^{(\g)}_{[0,L]}[\phi]\}$ as $\e\to 0$; for a more precise statement, proved in the case that $J$ and $v$ are both of finite range, see \cite{COP} . The nonlocal large deviation functional corresponding to Eq.(\ref{00.1}) has the form,
\be
\label{00.2}
\begin{split}
\mc E^{(\g)}_{[0,L]}[\phi] & = \int_0^L\!\rmd x\, F(\phi(x))+\frac14\int_0^L \!\rmd x \int_0^L
\! \rmd y\, J(x-y)\big[\phi(x)-\phi(y)\big]^2  \\
&\quad + \frac{\g}2 \int_0^L\!\rmd x \!\int_0^L\! \rmd y\, \phi(x) v\big(\g(x-y)\big) \phi(y) \;,
\end{split}
\ee
where $\phi(x)\in[-1,1]$, $F(t)=a(t)-\min_t a(t)$ and
\be 
\label{00.3}
a(t)=-\frac{\widehat J_0t^2}2+\frac1\b \Big(\frac{1+t}2\log\frac{1+t}2 + \frac{1-t}2\log\frac{1-t}2 \Big)\;, 
\ee
with $\widehat J_0=\int_{\bb R}\! \rmd x\, J(x)$. If $\b>\b_c=\widehat J_0^{-1}$, the local potential $F$ has a double well shape, with two degenerate minima located in $\pm m_\b$, where $m_\b$ is the positive solution to the self-consistency equation $m=\tanh(\b \widehat J_0 m)$. Therefore, for $\g=0$, the minimizers are the homogeneous profiles $\phi(x)\equiv m_\b$ or $\phi(x)\equiv -m_\b$. For $\g>0$ there is a competition between the short ranged part $\mc E^{(0)}_{[0,L]}[\phi]$, which favors the ``FM phase" $\phi\equiv m_\b$ or $\phi\equiv -m_\b$, and the long ranged part $\frac{\g}2 \int_0^L\!\rmd x \!\int_0^L\! \rmd y\, \phi(x) v\big(\g(x-y)\big) \phi(y)$, which favors the  ``paramagnetic phase" $\phi\equiv 0$. It is natural to expect that the competition favors the formation of magnetized structures of size intermediate between 1 (the range of $J(x)$) and $\g^{-1}$ (the range of $v(\g x)$) of alternating sign, so that the total magnetization is equal to zero; in this way the system appears to be uniformly magnetized $+m_\b$ or $-m_\b$, if seen on the scale of $J(x)$, while it appears to be paramagnetic, if seen on the scale of $v(\g x)$. Pictorially speaking, we expect that the system develops a ``froth" (or ``foam"), as first conjectured by Lebowitz and Penrose in their pioneering paper \cite{LP}. The main result of this paper is a characterization of the microscopic structure of the foam in the problem at hand and a proof that the profiles with energy close to the minimum of $\mc E^{(\g)}_{[0,L]}$ are ``essentially periodic" (in a sense to be clarified) and oscillate with an optimal modulation length of the order $\g^{-2/3}$.

%%%%%%%%%%%%%%%%%%%%%%%%%%%%%%%%%%%%%%%%%%%%%%%%%%%
\section{Main results and strategy of the proof}
\label{sec:2}
%%%%%%%%%%%%%%%%%%%%%%%%%%%%%%%%%%%%%%%%%%%%%%%%%%

Our goal is to characterize the shape of the ``quasi-minimizers" of $\mc E^{(\g)}_{[0,L]}[\phi]$ in the case that $\b>\b_c=\widehat J_0^{-1}$, that is when the function $F$ in Eq.(\ref{00.2}) has a double well shape. Loosely speaking, a quasi-minimizer  is a magnetization profile with energy ``sufficiently close" to the minimum; we shall clarify and quantify what we mean by that below. However, before doing this, we find convenient to introduce the sharp interface counterpart of the functional of interest, which was studied in \cite{GLL3} by RP, and to briefly review the key bounds that characterize its quasi-minimizers: these will justify and motivate the statement of the corresponding results in the non local functional studied in this paper. 

Let $\t$ be the surface energy associated to the short-ranged part of the functional, namely
\be
\label{2.1}
\tau = \inf_{\phi\in \mc M}\mc E ^{(0)}[\phi]\;,
\ee
where
\[
\mc E^{(0)}[\phi]=\int_{\bb R}\! \rmd x\, F(\phi(x))+\frac14\int_{\bb R}\! \rmd x \!\int_{\bb R}\! \rmd y\, J(x-y)\big[\phi(x)-\phi(y)\big]^2
\]
and 
\be 
\mc M = \left\{ \phi\in L^\io(\bb R;[-1,1])\colon \limsup_{x\to-\infty}\phi(x) <0\;, \;\;\liminf_{x\to+\io}\phi(x)>0\right\}\;.\label{2.inst}
\ee
We recall that the variational problem \eqref{2.1} has a minimizer which is unique up to translations \cite{DOPT1,DOPT2,Presutti}. More precisely, any minimizer has the form $q_z(x) = q(x-z)$, $z\in\bb R$, where $q(x)$ is a strictly monotone antisymmetric function, solution to the local mean field equation $q=\tanh\beta J*q$, and converging exponentially fast to $\pm m_\b$ as $x\to\pm\io$. In particular, $\tau = \mc E^{(0)}[q]>0$. Sometimes the profile $q(x)$ is pictorially called the ``instanton". 

As explained above, if $\g$ is very small, we expect that the profiles with minimal energy, or close to the minimal energy, consist of jumps from the negative to the positive phase separated by a distance that typically is much larger than the range of $J(x)$ and much smaller than the range of $v(\g x)$.  The transition from negative to positive or viceversa is performed so to make the short range part of the functional happy: therefore, we expect the quasi-minimizers to have a shape essentially equal to the instanton in the vicinity of the transition point and we expect the energy cost of the transition to be  essentially $\t$. The soliton tends to $\pm m_\b$ exponentially fast with the distance from the jump; if seen from ``far away", i.e., on scale much larger than 1, the soliton is seen as a sudden jump from $-m_\b$ to $m_\b$. Therefore, a natural effective functional that should describe well the energy cost of the quasi-minimizers, if seen on a scale intermediate between 1 and $\g^{-1}$, is the following ``sharp interface" functional, which was studied in \cite{GLL3},
\be
\label{1.2bis}
\lis{\mc E}^{(\g)}_{[0,L]}[\sigma]  = \t\mc N_L(\s)  + \frac{\g}2 \int_{0}^L\!\rmd x \!\int_0^L\! \rmd y\, \s(x) v\big(\g(x-y)\big) \s(y)\;,
\ee
where $\s:[0,L]\to \{\pm m_\b\}$, it has finite bounded variation and $\mc N_L[\s]$ is the number of jumps from $-m_\b$ to $m_\b$ or viceversa. Given such a $\s$, let $\{H_j\}_{j=1}^{\mc N_L[\s]}$ be the partition of $[0,L]$ consisting of the maximal intervals on which $\s$ is constant, which are separated among each other by the jump points; moreover,  we define $h_j=|H_j|$ to be the lengths of these intervals. Due to the exact locality of the surface tension term, the functional Eq.(\ref{1.2bis}) can be studied by RP methods, which imply the remarkable estimate \cite{GLL3},
\be 
\label{00.tris}
\lis{\mc E}^{(\g)}_{[0,L]}[\sigma] \ge \sum_{j=1}^{\mc N_L[\s]}h_j e(h_j)= Le(h^*)+\sum_{j=1}^{\mc N_L[\s]}h_j \big(e(h_j)-e(h^*)\big)\;,
\ee
where $e(h)$ is the energy per unit length in the thermodynamic limit of the periodic configuration consisting of intervals $H_j$ of constant length equal to $h$ and $e(h^*)=\min_{h\in\mathbb R}e(h)$. More precisely,  
\be 
\label{00.quater}
e(h)=\lim_{L\to\infty}\frac{\lis{\mc E}^{(\g)}_{[0,L]}[\sigma_h]}{L}\;,\qquad \s_h(x)=m_\b\, \sign(\sin(\p x/h))
\ee
and an explicit computation shows that 
\be
\label{eh}
e(h)=\frac{\t}{h}+\l m^2_\b\int\!\frac{\m(\rmd\a)}{\a}\, \bigg(1-\frac{\tanh(\a\g h/2)}{\a\g h/2}\bigg)\;, 
\ee
and
\be
\label{00.6}
\begin{split} 
e(h^*)& =\Big(\frac{9}{16}\t^2m_\b^2|v'(0^+)|\Big)^{1/3}\g^{2/3}\big(1+O(\g^{2/3})\big)\;,\\ h^* & =\Big(\frac{6\t}{|v'(0^+)| m_\b^2}\Big)^{1/3} \g^{-2/3} (1+O(\g^{2/3}))\;.
\end{split}
\ee
Combining the lower bound Eq.(\ref{00.tris}) with the upper bound 
\[
\min_\s \lis{\mc E}^{(\g)}_{[0,L]}[\sigma] \le \min_h\lis{\mc E}^{(\g)}_{[0,L]}[\sigma_h]
\]
gives, 
\be  
\label{00.7}
\lim_{L\to\io}\inf_{\s} \frac{\lis {\mc E}^{(\g)}_{[0,L]}[\s]}L= e(h^*)\;.
\ee
Moreover, the correction term $\sum_{j=1}^{\mc N_L[\s]}h_j \big(e(h_j)-e(h^*)\big)$ in Eq.(\ref{00.tris}) provides an explicit bound on the energy cost for picking a magnetization profile different from $\s^*:=\s_{h^*}$ or from one of its translates. In particular, it characterizes the quasi-minimizers of $\lis {\mc E}^{(\g)}_{[0,L]}[\s]$, in the sense that, if $\lis {\mc E}^{(\g)}_{[0,L]}[\s]-\lis {\mc E}^{(\g)}_{[0,L]}[\s^*]\le L\g^{\frac23+\e_0}$, for some $0<\e_0<\frac23$, then Eq.(\ref{00.tris}) easily implies that, for all $\e\in(0,\frac{\e_0}{2})$,
\be
\label{2.wr} 
L_\mathrm{wrong}:= \sum_{j=1}^{\mc N_L[\s]}h_j\,\c\big(|h_j-h^*|\ge h^*\g^{\e}\big)\le \const L\g^{\e_0-2\e}
\;.
\ee
Since the RP methods yielding Eq.(\ref{00.tris}) break down in the presence of a non local short ranged interaction, the idea is to study the quasi-minimizers of Eq.(\ref{00.2}) by first coarse graining the system on a scale intermediate between 1 and $\g^{-2/3}$, which is the expected oscillation scale of the quasi-minimizers, and to correspondingly reduce the study of $\mc E^{(\g)}_{[0,L]}$ to that of a sharp interface functional, analogous to $\lis{\mc E}^{(\g)}_{[0,L]}$; next, the effective sharp interface functional will be studied by the same RP methods of \cite{GLL3}. The combination of these two ingredients yields detailed informations and estimates on the shape of the quasi-minimizers; we are not able to prove the exact periodicity of the minimizers, because the coarse graining procedure produces error terms (which we explicitly bound in the following) that causes the optimal or quasi-optimal profiles to be close to a periodic profile but not necessarily periodic (at least, this is the most we can say). 

The main results on the non local functional Eq.(\ref{00.2}) are summarized in the following theorems.
\begin{theorem}
\label{thm00}
For any $\d\in(0,\frac13)$, there exists $C_\d>0$ such that, if $\g$ is small enough, 
\be 
\label{1.0}
\lim_{L\to\io}\inf_{\phi} \frac{{\mc E}^{(\g)}_{[0,L]}[\phi]}L= e(h^*)\big(1+r(\g)\big)\;,\qquad |r(\g)|\le C_\d\g^{\frac13-\d}\;,
\ee
where the infimum in the left-hand side runs over measurable functions such that $\phi(x)\in[-1,1]$ for any $x\in[0,L]$. 
\end{theorem}
In order to state our results on the shape of the quasi-minimizers we need a few more definitions. Let $E_{0}^{(\g)}(L)=\inf_{\phi}{\mc E}^{(\g)}_{[0,L]}[\phi]$ be the ground state energy associated to the functional ${\mc E}^{(\g)}_{[0,L]}$ at finite $\g$ and finite $L$. Moreover, given $0<\d_0<\frac13$, let ${\mc P}_{\d_0}=\{Q_i\}_{i=1}^{M_L}$ be the partition of $[0,L]$ into $M_L$ intervals of length $\a_L(\d_0)\g^{-\d_0}$, where $\a_L(\d_0)=\{\inf \a\ge 1\colon \frac{L}{\a}\g^{\d_0}\in{\mathbb N}\}$; in the following, we shall refer to these intervals of  length $\sim\g^{-\d_0}$ as ``blocks". We define $\psi_{\phi}^{(\d_0)}$ to be the coarse version of $\phi$ on ${\mc P}_{\d_0}$, that is, $\psi_\phi^{(\d_0)}$ is the function, measurable with respect to $\mc P_{\d_0}$, whose value at $x\in Q_i$ is equal to $\media{\phi}_{Q_i}$, where $Q_i\in\mc P_{\d_0}$ and $\media{\phi}_{Q_i}=|Q_i|^{-1}\int_{Q_i}\!\rmd x\, \phi(x)$. Given $\phi$, we shall say that the block $Q_i\in\mc P_{\d_0}$ is of type $+$ (resp.\ $-$) if $\media{\phi}_{Q_i}\ge \frac9{10}m_\b$ (resp.\ $\media{\phi}_{Q_i}\le -\frac9{10}m_\b$); we shall say that it is of type $0$ if $|\media{\phi}_{Q_i}|< \frac9{10}m_\b$. The coarse version $\psi_\phi^{(\d_0)}$ of any function $\phi$ with energy sufficiently close to the minimum consists of long sequences of blocks of type $+$, of total length $h^*(1+o(1))$, followed by long sequences of blocks of type $-$, of total length $h^*(1+o(1))$, except for an infinitesimal fraction of $[0,L]$, on which $\psi_{\phi}^{(\d_0)}$ looks ``wrong" (i.e., has long sequences of blocks of type $0$ or sequences of $+$ or of $-$ of length different from $h^*$). This statement is made more precise by the following theorem. 
\begin{theorem}
\label{thm01} 
Given $0<\d_0<\e_0<\frac13$ and $L_0>0$, there exists a positive constant $\g_0$ such that, if $0< \g\le \g_0$ and $L\ge L_0\g^{-1}$, then:
\\
For any $\phi$ for which  
\be
\label{1.1} 
{\mc E}^{(\g)}_{[0,L]}[\phi]\le E_0^{(\g)}(L)+L\g^{\frac23+\e_0}\;,
\ee 
there exists a set $G_\phi\subseteq[0,L]$, measurable with respect to $\mc P_{\d_0}$, such that $|G_\phi|\ge L(1-\frac{16}{\t}\g^{\frac{\e_0}2})$, which is a disjoint union of intervals $\L_k$, all of length larger than $\g^{-\frac23-\frac{\e_0}{2}}$. The blocks $Q_i\in\mc P_{\d_0}$ contained in $G_\phi$ can be grouped into maximal connected sequences of blocks of constant type, either $+$ or $-$, to be called $I_j$, $j=1,\ldots, \mc N_L^\phi$. On each $\L_k$, the intervals $I_j$ have alternating sign and are separated among each other by at most one block of type $0$. Moreover, 
\be 
\label{1.2}
\sum_{j=1}^{\mc N_L^\phi}\bigg[|I_j|(e(|I_j|)-e(h^*))+\frac{F(0)}{m^2_\b} \int_{{\rm int}\,I_j}\!\rmd x\, (|\psi_{\phi}^{(\d_0)}(x)|-m_\b)^2\bigg]\le 10 L\g^{\frac23+\e_0}\;,
\ee
where ${\rm int}\,I_j$ indicates the subset of $I_j$ obtained from $I_j$ by depriving it of its first and last block in the sequence it consists of.
\end{theorem}
\begin{remark}\rm
\label{rem:p1}
The condition $L\ge L_0\g^{-1}$ is not sharp and it can be easily weakened to $L\ge \const\,\g^{-\frac23-\r}$, with $\r>0$, at the price of adding some extra conditions on $\e_0$ and $\d_0$. For such long intervals, the specific choice of boundary conditions that we made (``open boundary conditions") is irrelevant as far as the validity of Theorem \ref{thm01} is concerned, that is, the errors in energy that we make in changing from open to ``$\phi_\mathrm{out}$ boundary conditions" (which are of  order 1) can be absorbed into the error of order $L\g^{\frac23+\e_0}$ appearing in the statement of the theorem. Here, given any ``boundary condition" $\phi_\mathrm{out}$, i.e., any arbitrarily prefixed function $\phi_\mathrm{out}:\mathbb R\to[-,1,1]$, the functional $\mc E^{(\g);\phi_\mathrm{out}}_{\L}[\phi]$ with $\phi_\mathrm{out}$ boundary conditions on $\L=[0,L]$ is defined as 
\[
\begin{split}
& {\mc E}^{(\g);\phi_\mathrm{out}}_{\L}[\phi] = {\mc E}^{(\g)}_{\L}[\phi] \\ & \quad\qquad +\int_{\L}\!\rmd x \int_{\L^c}\! \rmd y\,\Big[\frac12 J(x-y)\big(\phi(x)-\phi_\mathrm{out}(y)\big)^2+ \g\phi(x) v\big(\g(x-y)\big)\phi_\mathrm{out}(y)\Big]\;.
\end{split}
\]
The cases in which $\phi_\mathrm{out}$ is equal to $m_\b$, or $-m_\b$, or to the periodic extension of $\phi:\L\to[-1,1]$ to $\mathbb R$, or to the Neumann extension of $\phi:\L\to[-1,1]$ to $\mathbb R$ (i.e., the function obtained from $\phi$ by repeated reflections about the endpoints of $\L$), are special and are refereed to as $+$ boundary conditions (b.c.), or $-$ b.c., or periodic b.c., or Neumann b.c., respectively. Let us note that these four special boundary conditions ``are better than others", in particular they are better than open b.c.: by this we mean that in the presence of such boundary conditions there are no $O(1)$ error terms entering the estimates due to the boundary conditions. This makes possible to study the limiting behavior of the functional $\mc E^{(\g);\mathrm{per}}_{[0,L]}$ with (say) periodic boundary conditions as $\g\to 0$ on intervals of length of the order $\g^{-2/3}$, that is the same scale as the optimal oscillation length $h^*$. This is an interesting case by itself, which is discussed in Subsection \ref{sec3:gamma}.
\end{remark}

A useful corollary of Theorem \ref{thm01}, and in particular of Eq.(\ref{1.2}), is the following: define the sets 
\[
\begin{split}
X_1 & =\bigcup_{j=1}^{\mc N_L^\phi}\big\{x\in {\rm int}\, I_j\colon \big||\psi_\phi^{(\d_0)}(x)|-m_\b\big|\ge \g^{\e}\big\}\;,\\ X_2 & =\bigcup_{j=1}^{\mc N_L^\phi}\big\{I_j\colon \big||I_{j}|-h^*\big|\ge h^*\g^{\e'}\big\}\;,
\end{split}
\]
where
$\e\in(0,\frac13+\frac{\e_0}2)$ and $\e'\in(0,\frac{\e_0}{2})$. Note that the sets $X_1$ and $X_2$ can be thought of as the intervals where ``things go wrong", either because $\psi_\phi^{(\d_0)}$ is substantially different from $\pm m_\b$, or because $|I_j|$ is substantially different from the expected optimal length $h^*$. 
\begin{corollary}
\label{cor1}
Under the same assumptions of Theorem \ref{thm01}, if $\phi$ fulfills Eq.(\ref{1.1}), then 
\be 
\label{1.3}
|X_1|\le C L\g^{\frac23+\e_0-2\e}\;,\qquad |X_2|\le CL\g^{\e_0-2\e'}\;,
\ee
for a suitable constant $C>0$. 
\end{corollary}
Theorem \ref{thm01} and its corollary characterize the quasi-minimizers of $\mc E^{(\g)}_{[0,L]}$ for $\g$ small, asymptotically as $L\to\infty$. In particular, the two estimates in Eq.(\ref{1.3}) are the analogues of Eq.(\ref{2.wr}). The proofs of these claims are based on a coarse  graining procedure that maps every measurable function $\phi$ into a piecewise constant function $\s_\phi$ such that $|\s_\phi|\ge m_\b(1-o(1))$. An essential condition on the coarse grain procedure is that it induces  a small change in the long range contributions to the energy. This is realized by conserving the averages  on suitably long blocks $B_i$, where we require $\media{\phi}_{B_i}=\media{\sigma_\phi}_{B_i}$. As a consequence we are forced to solve constrained variational problems (in the ``canonical ensemble'') on such blocks and to face delicate finite size effects. Among them the arising of a {\it critical droplet size} which is treated by means of arguments similar to those employed in \cite{CCELM2} in higher dimension. As a result of this construction, the original functional is bounded from below in terms of a new functional $\widetilde{\mc E}^{(\g)}_{[0,L]}$, acting on the space of the $\s_\phi$'s, which is simpler than the original one, because it has a local surface tension term and, therefore, can be studied by the reflection positivity methods of \cite{GLL1,GLL2,GLL3}. 

\medskip
The rest of the paper is organized as follows. In Section \ref{sec:3} we describe the coarse graining procedure that maps every profile $\phi(x)$ into a piecewise constant function $\s_\phi(x)$ and, correspondingly, bounds the original functional $\mc E^{(\g)}_\L$ from below in terms of a simplified functional $\widetilde{\mc E}^{(\g)}_{\L}$ for $\s_\phi$, which can be studied by RP methods; the main results of this section is summarized in Proposition \ref{prop:0}. In Section \ref{sec:4} we prove Theorems \ref{thm00} and \ref{thm01}, by using Proposition \ref{prop:0}. In Section \ref{sec:5} we prove Proposition \ref{prop:0}. Finally, in Section \ref{sec:conc} we draw the conclusions and discuss some open problems. Some technical aspects of the proofs are deferred to the appendices.

%%%%%%%%%%%%%%%%%%%%%%%%%%%%%%%%%%%%%%%%%%%%%%%%%%
\section{The coarse graining procedure}
\label{sec:3}
%%%%%%%%%%%%%%%%%%%%%%%%%%%%%%%%%%%%%%%%%%%%%%%%%%

In this section, we describe the coarse graining procedure that maps every profile $\phi(x)$, $x\in [0,L]$, into a piecewise constant function $\s_\phi(x)$ such that $|\sigma_\phi(x)|\ge \lis m:=m_\b-\k \g^{\d/2}$, where $\k$ is a suitable positive constant, to be fixed below. We denote by $K_{[0,L]}$ the space of such functions. Given $\s\in K_{[0,L]}$, we denote by $H_j$ the maximal intervals on which $\s$ has constant sign and by $h_j=|H_j|$ their lengths, with $j=1,\ldots,N_L^\s$. We shall say that an interval $H_j$ is of type + or $-$, depending on whether $\s$ is positive or negative on it; in this sense, $\s$ induces a partition of $[0,L]$ consisting of intervals of alternating type, on which $\s$ is correspondingly positive or negative. The relevance of the map $\phi\to\s_\phi$ relies on the fact that the energy of $\phi$ can be bounded from below in terms of the energy of $\sigma_\phi(x)$, which is computed by using a modified functional $\widetilde{\mc E}^{(\g)}_{[0,L]}[\sigma_\phi]$; moreover, the minimizers of the latter can be estimated by using the methods of \cite{GLL1,GLL2,GLL3}. The result is summarized in the following proposition, which is proved in Section \ref{sec:5}.
\begin{proposition}
\label{prop:0}
Let $\d\in(0,\frac13)$ and $L_0>0$. There exists $\bar\g_0=\bar\g_0(\d,L_0)$ such that, if $0<\g\le \bar\g_0$ and $L\ge L_0\g^{-1}$, then for any measurable function $\phi:[0,L]\to[-1,1]$ there exists a piecewise constant function $\s_\phi\in K_{[0,L]}$, such that 
\be
\label{sp}
\mc E^{(\g)}_{[0,L]}[\phi] \ge \widetilde{\mc E}^{(\g)}_{[0,L]}[\s_\phi]+O(\g^{1-\d})L\;,
\ee 
where 
\be
\label{1.2ybis}
\begin{split}
\widetilde{\mc E}^{(\g)}_{[0,L]}[\sigma] & = \frac{F(0)}{2m_\b^2} \int_{0}^L\!\rmd x\, (|\s(x)|-m_\b)^2 \\ & \quad + \frac{\t}2 \int_0^L\!\rmd x\, \Big|\frac{\rmd}{\rmd x} \frac{\s(x)}{|\s(x)|}\Big| + \frac{\g}2 \int_{0}^L\!\rmd x \!\int_0^L\! \rmd y\, \s(x) v\big(\g(x-y)\big) \s(y)\;.
\end{split}
\ee
Moreover, for any $L>0$ and any $\s\in K_{[0,L]}$, 
\be 
\label{c000}
\widetilde{\mc E}^{(\g)}_{[0,L]}[\s]\ge L e(h^*)+\frac12\sum_{j=1}^{N_L^\s}h_j(e(h_j)-e(h^*))+ \frac{F(0)}{4m_\b^2}\int_0^L\!\rmd x\, (|\s(x)|-m_\b)^2+O(\g^{\frac43})L\;.
\ee
\end{proposition}
\begin{remark}\rm
\label{rem:p2}
The function $\widetilde F(\s)= \frac{F(0)}{2m_\b^2}(|\s|-m_\b)^2$ appearing in $\widetilde{\mc E}^{(0)}_{[0,L]}[\s]$ is even and its restriction to $\s\ge 0$ is convex. Moreover, it is such that $\widetilde F(t)\le \frac12F(t)$, a property that will be used in the proof of Proposition \ref{prop:0} and is proven in Appendix \ref{sec:A}.
\end{remark}
Of course, in order to make the statement of Proposition \ref{prop:0} more explicit, we need to explain how the reference profile $\s_\phi$ is defined, which is done in Subsection \ref{sec3:partition} below. However, before doing that, let us add a few more remarks about the connection between Proposition \ref{prop:0} and the notion of $\G$-convergence.

\subsection{On the relation with $\G$-convergence}
\label{sec3:gamma}
Proposition \ref{prop:0}, which is the key technical result behind the proofs of our main results announced in Section \ref{sec:2}, is in many respects stronger than Theorems \ref{thm00} and \ref{thm01}. In fact, Proposition \ref{prop:0} provides us with detailed informations about the ``excited states", rather than just the minimizers or the quasi-minimizers, of our variational problem. Consider the space $\mc K_{[0,L]}$ of functions on $[0,L]$ that assume only values $\pm m_\b$ and note that if $u\in \mc K_{[0,L]}$, the functional $\widetilde{\mc E}^{(\gamma)}_{[0,L]}[u]$ is the same as the functional $\lis{\mc E}^{(\g)}_{[0,L]}[u]$ discussed in Section \ref{sec:2}, for which the energy of any (not necessarily minimal) configuration can be efficiently estimated by RP methods. If $\phi$ is close to a profile $u\in \mc K_{[0,L]}$ (here ``close" means that  both $\sum_{j}h_j(e(h_j)-e(h^*))$ and $\int_0^L\!\rmd x\,(|\phi|-m_\b)^2$ are small as $\g\to 0$), then Proposition \ref{prop:0} tells us that $\mc E^{(\g)}_{[0,L]}[\phi]$ is bounded from below by $\lis{\mc E}^{(\g)}_{[0,L]}[u]$ plus small error terms as $\g\to 0$. An inequality in the opposite direction is valid, too: if $u\in \mc K_{[0,L]}$ is ``reasonable" (i.e., if the distance between its jump points is larger than $O(\log^2\g)$) then one can find a smooth profile $\phi_u$ (which is obtained from $u$ by replacing the sharp interfaces by cut-offed instantonic profiles) such that $\lis{\mc E}^{(\g)}_{[0,L]}[u]$ is bounded from below by ${\mc E}^{(\g)}_{[0,L]}[\phi_u]$, up to small errors as $\g\to 0$. In other words, our functional of interest is bounded from above and below by $\lis{\mc E}^{(\g)}_{[0,L]}[u]$ up to small error terms as $\g\to 0$: in this sense Proposition \ref{prop:0} can be thought of as a quantitative version of De Giorgi's $\G$-convergence \cite{Braides, DalMaso} in the ``thermodynamic limit" (i.e., with error terms scaling proportionally to the length of the interval $L$).

\smallskip
We recall that the sequence $\mc F^{(\g)}$ of functionals on a metric function space $\mc K$ is said to $\Gamma$-converge to $\mc F$ as $\g\to0$ if
\begin{enumerate}
\item \textit{($\Gamma$-liminf)}. For each $u\in\mc K$ and each sequence $u_\g$ converging to $u$ in $\mc K$, it holds that  $\liminf_{\g\to 0}\mc F^{(\g)}[u_\g]\ge \mc F[u]$.
\item \textit{($\Gamma$-limsup)}. For each $u\in\mc K$ there exists a sequence $u_\g$ converging to $u$ in $\mc K$ such that  $\limsup_{\g\to 0}\mc F^{(\g)}[u_\g]= \mc F[u]$.
\end{enumerate}
Moreover, see \cite[Theorem 1.21]{Braides}, if the sequence is equicoercive (i.e., if any sequence $u_\g$ such that $\limsup_{\g\to 0}\mc F^{(\g)}[u_\g]<+\infty$ is precompact) then $\Gamma$-convergence  implies the convergence of the minimizers.

If we insist in taking $\g\to 0$ rather than keeping it finite with explicit error terms, we can translate part of the results of Proposition \ref{prop:0} into a $\G$-convergence result, for instance in the case that $L$ is chosen proportionally to $\g^{-\frac23}$ (which is a possible choice for $L$ if, say, periodic boundary conditions are chosen, see  Remark  \ref{rem:p1}). Let us explain this in some more detail: we choose periodic boundary conditions and $L=L_0\g^{-2/3}$, with fixed $L_0>1$, and we adopt the rescaled variable $r = \g^{2/3}x$. The energy, as a function of the $L_0$-periodic profiles $\psi(r) = \phi(\g^{-2/3}r)$, is easily seen to be given by the following functional,
\[
\mc F^{(\g)}[\psi] = \mc F^{(\g)}_0[\psi] + \mc F^{(\g)}_1[\psi]\;, 
\]
where
\[
\begin{split}
\mc F^{(\g)}_0[\psi] & = \g^{-2/3} \int_0^{L_0}\!\rmd r\, F(\psi(r))+\frac{\g^{-2/3}}4 \int_0^{L_0} \!\rmd r \int_0^{L_0}\! \rmd r'\, J^\mathrm{per}_\g(r,r')\big[\psi(r)-\psi(r')\big]^2\;,  \\ \mc F^{(\g)}_1[\psi] & = \frac{\g^{-2/3}}2\int_0^{L_0} \!\rmd r \!\int_0^{L_0} \! \rmd r'\, \psi(r) v^\mathrm{per}_\g(r,r') \psi(r')\;,
\end{split}
\]
with
\[
\begin{split}
J^\mathrm{per}_\g(r,r') & = \sum_{n\in \bb Z} \g^{-2/3}J\big(\g^{-2/3}(r-r'-nL)\big)\;,  \\ v^\mathrm{per}_\g(r,r') & = \sum_{n\in \bb Z} \g^{1/3}v\big(\g^{1/3}(r-r'-nL)\big)\;.
\end{split}
\]
Then, it is easy to show that  the sequence $\mc F^{(\g)}$ on $L^1([0,L_0])$ is equicoercive and $\G$-converges to 
\[
\mc F[u] = \begin{cases} {\displaystyle \frac{\tau}{2m_\b} \int_0^{L_0}\!\rmd r\, |u'(r)| + \l\media{\a}\int_0^{L_0}\!\rmd r\, |(-\Delta)^{-1/2}u(r)|^2} & \mathrm{if}\quad \int_0^{L_0} u(r)dr=0\;, \\ +\infty & \mathrm{otherwise}\;, \end{cases}
\]
where $(-\Delta)^{-1}$ is the inverse of the Laplacian on $[0,L_0]$ with periodic boundary condition. The minimizers of this functional have been studied in \cite{RW} where it is proved that they are periodic. We do not belabor the details of this statement, which is a simple consequence of an analogous statement for the short ranged part of the functional, see e.g. \cite[Section 7.1.7]{Presutti}, and a straightforward bound on the difference between $\mc F^{(\g)}_1[\psi]$ and $\frac 12 \g^{-2/3}L m^2 \widehat v_0+  \l\media{\a}  \|(-\Delta)^{-1/2}(\psi-m)\|_2^2$.

Of course, as in any $\G$-convergence result, the form of the limit depends on the rescaling chosen both for the lengths and the energies. The special rescaling chosen above is interesting and natural, because in the limit both the short and the long range interaction terms survive and compete on equal footings, as in the original finite-$\g$ functional of interest. Still, it may be interesting to investigate in a more detailed way the possibility of defining more precisely a notion of $\G$-convergence in infinite (or at least larger than $L_0\g^{2/3}$) volume. We hope to come back to this issue in a future publication.
\\
Let us now come back to the description of the map $\phi\to \s_\phi$.

\subsection{Partitioning the big interval.} 
\label{sec3:partition}
The replacement of $\phi$ into $\s_\phi$ is a local procedure, defined in each single element $B_i$ of a suitable partition of $[0,L]$ and depending on the average of $\phi$ on each such element. Therefore, the first thing that we need to explain is how to define the partition $\{B_i\}$, {\it which depends on the shape of $\phi$} itself. We start from a regular partition and then modify some of the intervals, adapting them to the shape of $\phi$. Let us fix $\d\in(0,\frac13)$ and consider the partition $\mc P_\d$ of $[0,L]$ into intervals of constant length $\ell_+=\a_L(\d)\g^{-\d}$, which was defined after Theorem \ref{thm00}. Given $\phi$ on $[0,L]$, let us label each block $Q_i$ in $\mc P_\d$ by the value of its internal energy $\mc E^{(0)}_{Q_i}[\phi]$. We start by selecting the blocks whose internal energy is not bigger than $2\t$. Not surprisingly, if $Q_i$ has energy that does not exceed this cut-off value, then it is possible to find a long segment in $Q_i$ where $\phi$ stays close either to $+m_\b$ or to $-m_\b$; moreover, this segment can be chosen to stay sufficiently far from the boundary of $Q_i$. For a precise statement, see the following lemma (and see Appendix \ref{sec:B} for its proof). 
\begin{lemma}
\label{lem:1}
Given a block $Q_i\in\mc P_\d$, let us partition it into a sequence of small blocks of size $\ell_-\ll1$, with $\ell_-$ a small $O(1)$ number, independent of $\g$. Given a small block $b_j$, let us denote by $\media{\phi}_{b_j}$ the average of $\phi$ over $b_j$. If $\mc E^{(0)}_{Q_i}[\phi]\le 2\t$, then for any $0<\r<\d/2$ there exists $\o\in\{\pm1\}$ such that  it is possible to find a sequence $\mc S_\o$ of $M$ contiguous small blocks with the following properties:
\\ \noindent
1) if $b_j\in\mc S_\o$, then $|\media{\phi}_{b_j}-\o m_\b|\le \g^{\r}$;
\\ \noindent
2) the total length $M\ell_-$ of $\mc S_\o$ is larger than $\bar C\g^{-(\d-2\r)}\gg 1$, for a suitable constant $\bar C$ (possibly depending on $\ell_-$);
\\ \noindent
3) the distance of each block in $\mc S_\o$ from the boundary of $Q_i$ is larger than $\ell_+/4$.
\end{lemma}
\begin{figure}\centering
\includegraphics[scale=.4,angle=0,draft=false]{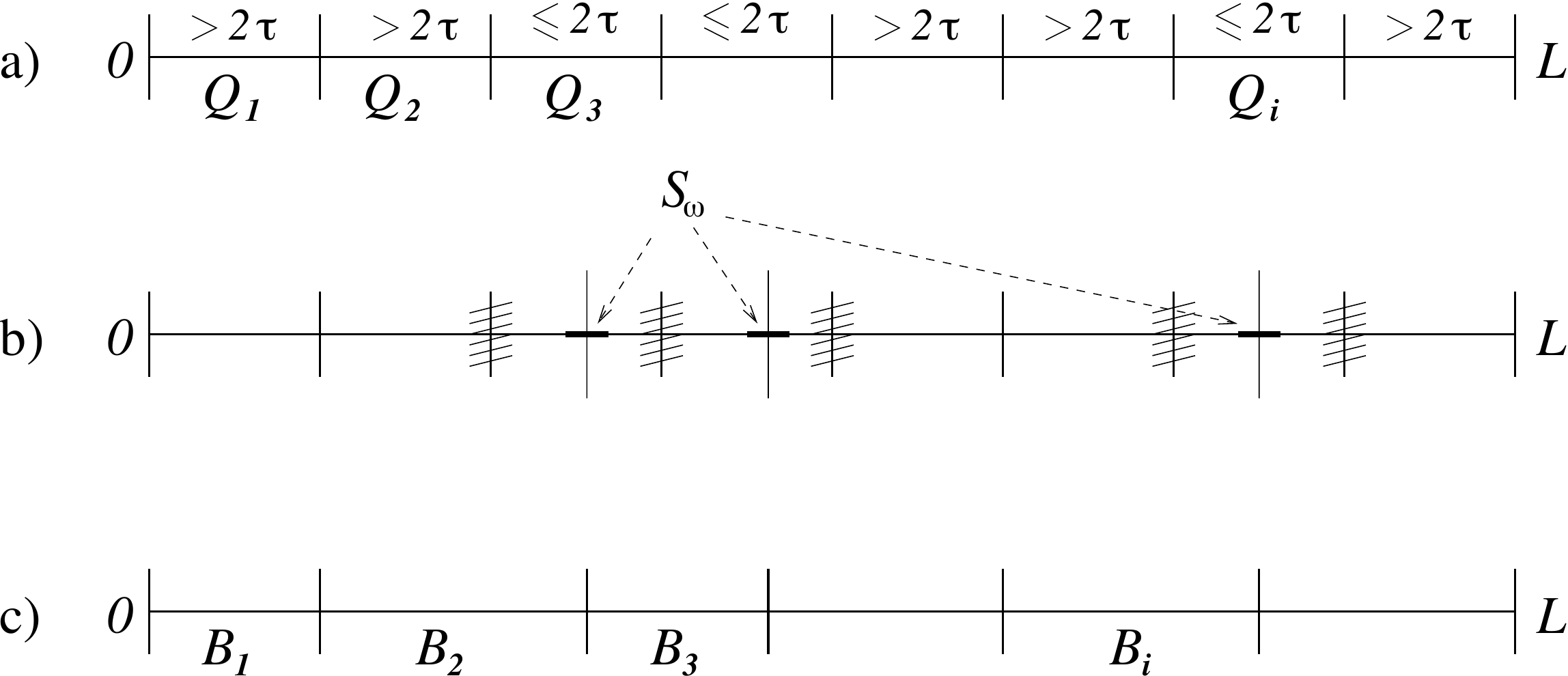}
\caption{The partition $\{B_i\}$ in c) is obtained  from the partition $\{Q_i\}$ in a) by removing the boundary lines of each block $Q_i$ with internal energy not bigger than $2\t$ and  adding a new boundary line in the middle of the corresponding segment $\mc S_\o$, as shown in b).}
\label{fig:1}
\end{figure}
Now, we modify the original partition of $[0,L]$ in the following way. For each block $Q_i$ with internal energy not bigger than $2\t$, we remove its boundary lines and draw a new ``boundary line'' in the middle of the segment $\mc S_\o$ defined in Lemma \ref{lem:1}, see Fig.\ \ref{fig:1}. We end up with a sequence of segments $V_k$ of $[0,L]$ delimited by two boundary lines, each of which ``well in the middle'' of a region where $\phi$ is essentially constant. Each $V_k$ has the property of being partitioned into new blocks $B_i$ of size $\sim \ell_+$ (i.e., of size $\frac 12 \ell_+\le \ell\le \frac52\ell_+$) and, either $V_k$ consits of a single block $B_i$, to be called a ``good block" (note that by construction each good block -- with the possible exception of a ``boundary" good block, i.e., a good block that is adjacent to the boundary of $[0,L]$ -- has $(\o,\o')$ boundary conditions, where $\o$ is the sign of $\phi$ at the left boundary and $\o'$ is the sign of $\phi$ at the right boundary), or each of its blocks $B_i$ has an internal energy $> 2\t$ (in which case we will say that $V_k$ consists of a collection of bad blocks). These new blocks form the desired partition $\BBB_\phi=\{B_i\}$ of $[0,L]$. We will further denote by $\BBB_\phi^g$ the set of good blocks in $\BBB_\phi$ and by $\BBB^b_\phi$ the set of bad blocks. 

\subsection{Replacing $\phi$ in each block of the partition.}
\label{subsec3.2} 
We now need to explain how to replace $\phi$ by $\s_\phi$ within each element of $\BBB_\phi$. Given $B_i\in\BBB_\phi$, we shall assume without loss of generality that $B_i=[0,\ell_i]$, with $\ell_i:=|B_i|$, and let $m_i=\media{\phi}_{B_i}$. We also introduce a tolerance $\z:=c_0 \g^{\d}\log^2\g$, with $c_0>0$ a suitable constant to be fixed below, which will be used to distinguish the blocks where the average of $\phi$ is larger than $m_\b-\z$ in absolute value, from those where it is smaller. The replacement procedure within $B_i$ depends on whether $B_i$ is good or bad, as explained in the following. For an example,
see Fig.\ 2.
\begin{figure}\centering
\includegraphics[scale=.4,angle=0,draft=false]{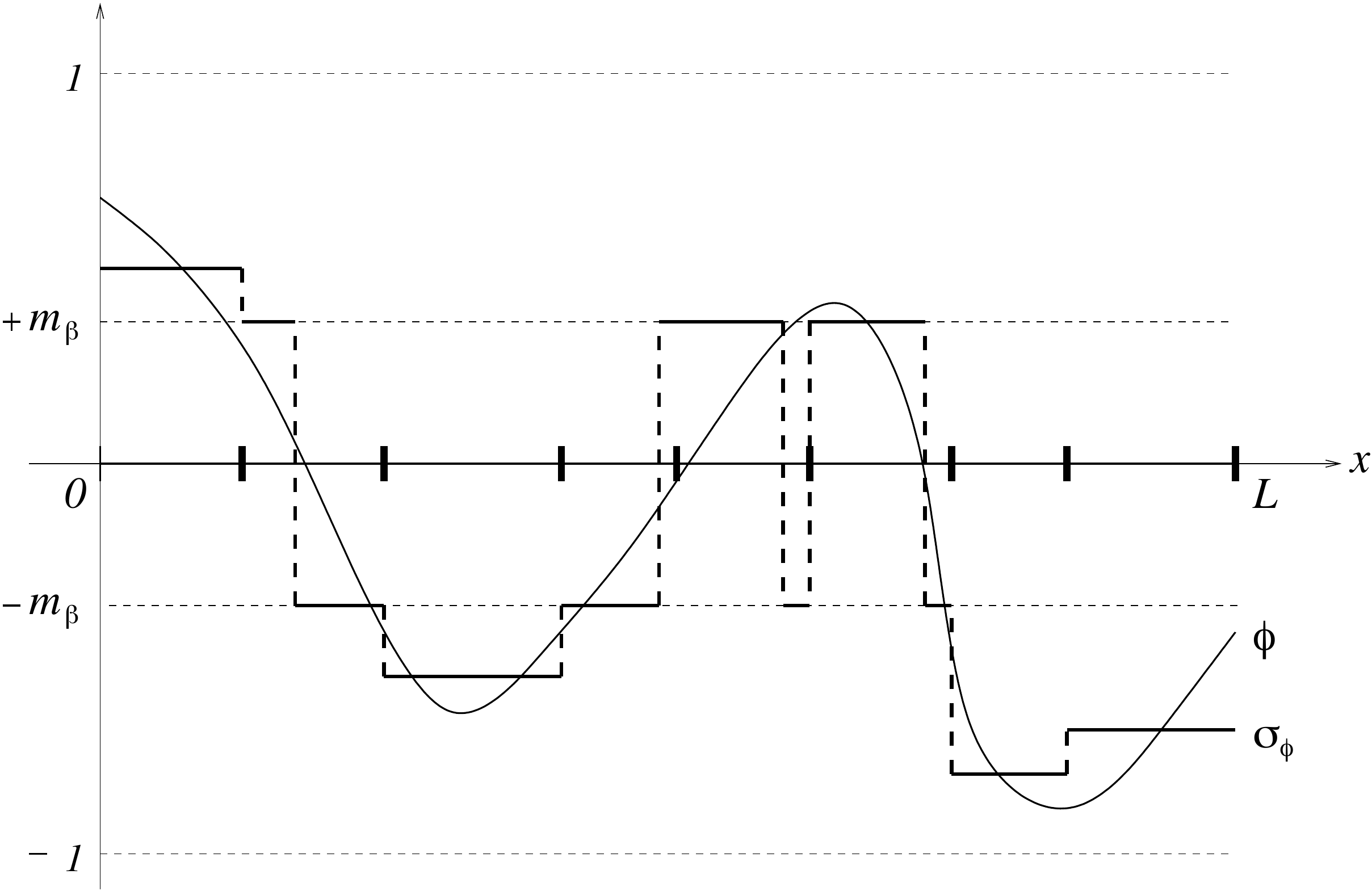}
\caption{An example of the replacement procedure $\phi\to \s_\phi$.  In the picture, the smooth curve is $\phi$ and the piecewise constant function is $\s_\phi$. The notches on the $x$ axis indicate the endpoints of the blocks $B_i$. The rationale behind the example is that in the blocks $B_i$ where $|\media{\phi}_{B_i}|>m_\b$, we replace $\phi$ by $\s_\phi:=\media{\phi}_{B_i}$; in the blocks where $|\media{\phi}_{B_i}|<m_\b$, we replace $\phi$ by a piecewise constant function $\s_\phi$ such that $|\s_\phi|=m_\b$ with at most two jump points. Note that in general, there are a few exceptional cases, corresponding to situations where $\media{\phi}_{B_i}$ is very close to $m_\b$ or $-m_\b$, for which the replacement procedure is more complicated than the one illustrated in the figure (see e.g. case (2a) below with $m_i>m_\b-\z$).}
\label{fig:2}
\end{figure}
\begin{enumerate}
\item In the case that $B_i\in\BBB^b_\phi$, the replacement procedure is very simple: if $|m_i|\ge m_\b-\z$, then $\s_\phi=m_i$ on $B_i$.
If $|m_i|<m_\b-\z$, then $\s_\phi=m_\b$ on $[0,\x]$ and $=-m_\b$ on $[\x,\ell_i]$, with $\x$ fixed in such a way that $\media{\s_\phi}_{B_i}=m_i$. We remark that the choice to have $\s_\phi$ positive to the left and negative to the right is arbitrary and not necessary. 
\item In the case that $B_i\in\BBB^g_\phi$, the replacement procedure is more elaborated and depends on the specific boundary conditions associated with $B_i$.
\begin{enumerate}
\item Suppose that $B_i$ is not a boundary block and the boundary conditions are $(\o,-\o)$. If $|m|\le m_\b-\z$, then $\s_\phi=\o m_\b$ on $[0,\x]$ and $=-\o m_\b$ on $[\x,\ell_i]$, with $\x$ fixed in such a way that $\media{\s_\phi}_{B_i}=m_i$. In the opposite case, let us assume $m_i>m_\b-\zeta$ (the occurrence $m_i<-m_\beta+\zeta$ can be treated analogously). We decompose $B_i$ as the union of three intervals, $B_i=\mathcal{I}\cup\mathcal{I'}\cup\mathcal{I''}$, with $|\mathcal{I'}|= |\mathcal{I''}|=\frac 12 \log^2\ell_i$ and $\mathcal{I}'$ [resp.\ $\mathcal{I''}$] on the side where the boundary condition is positive [resp.\ negative], and set
\be 
\label{15}
\sigma_\phi(x) = \begin{cases} m_i\ell_i(\ell_i-\log^2\ell_i)^{-1} & {\rm if}\quad x\in \mathcal{I}\;,\\ m_\b & {\rm if}\quad x\in \mathcal{I'} \;,\\  -m_\b & {\rm if}\quad x\in \mathcal{I''}\;.\end{cases}
\ee
Note that $\media{\s_\phi}_{B_i}=m_i$. Moreover, it will be proved below that $\s_\phi(x)\le 1$ for $x\in\mc I$, as it should.
\item Suppose that $B_i$ is not a boundary block and the boundary conditions are $(\o,\o)$. Without loss of generality, let us assume that $(\o,\o)=(-,-)$ (of course, the case of $(+,+)$ boundary conditions is treated similarly). If $-1\le m_i \le - m_\b + \frac{11}{10}C_*\ell_i^{-1/2}$, with $C_*=\sqrt{\frac{5\t}{F''(m_\b)}}$, then $\s_\phi(x)\equiv m_i$ on $B_i$. If $ - m_\b + \frac{11}{10}C_*\ell_i^{-1/2}<m_i<m_\b-\frac{11}{10}C_*\ell_i^{-1/2}$, then $\s_\phi(x)=-m_\b$ on $[0,\x]\cup[\ell_i-\x,\ell_i]$ and $=m_\b$ on $[\x,\ell_i-\x]$, with $\x$ fixed in such a way that $\media{\s_\phi}_{B_i}=m_i$. If $m_i > m_\b - \frac{11}{10}C_*\ell_i^{-1/2}$, then, proceeding as in the previous item, we decompose $B_i$ as the union of three intervals, $B_i=\mathcal{I}\cup\mathcal{I'}\cup\mathcal{I''}$, with $\mathcal{I'}=[0,\frac 12 \log^2\ell_i]$ and $\mathcal{I}''=[\ell_i-\frac 12 \log^2\ell_i,\ell_i]$, and set
\[
\sigma_\phi(x) = \begin{cases} (m_i\ell_i + m_\b \log^2\ell_i)(\ell_i-\log^2\ell_i)^{-1} & {\rm if}\quad x\in \mathcal{I}\;,\\ - m_\b & {\rm if}\quad x\in \mathcal{I'}\cup \mathcal{I''}\;.\end{cases}
\]
Note that also in this case $\media{\s_\phi}_{B_i}=m_i$. 
\item Suppose that $B_i$ is a boundary block. Consider, without loss of generality, the case that $B_i=[0,\ell_i]$ is adjacent to the left boundary of $[0,L]$, and the boundary condition at the right end of $B_i$ is $\o=-$. Then: if $-1\le m_i \le - m_\b + \frac{11}{10}C_*\ell_i^{-1/2}$, then $\s_\phi(x)\equiv m_i$ on $B_i$; if $ - m_\b +\frac{11}{10} C_*\ell_i^{-1/2}<m_i<m_\b-\frac{11}{10}C_*\ell_i^{-1/2}$, then $\s_\phi(x)=-m_\b$ on $[\ell_i-\x,\ell_i]$ and $=m_\b$ on $[0,\ell_i-\x]$, with $\x$ fixed in such a way that $\media{\s_\phi}_{B_i}=m_i$; if $m_i > m_\b - \frac{11}{10}C_*\ell_i^{-1/2}$, then
\[
\sigma_\phi(x) = \begin{cases} (2m_i\ell_i + m_\b \log^22\ell_i)(2\ell_i-\log^22\ell_i)^{-1} & {\rm if}\quad x\in [0,\ell_i-
\frac12\log^22\ell_i]\;,\\ -m_\b & {\rm if}\quad x\in [\ell_i-\frac12\log^22\ell_i,\ell_i]\;.\end{cases}
\]
Note that also in this case $\media{\s_\phi}_{B_i}=m_i$. 
\end{enumerate}
\end{enumerate}
This concludes the description of the map $\phi\to\s_\phi$ in the case of open boundary conditions. The same definitions are also valid in the case of generic $\phi_\mathrm{out}$ boundary conditions, with a couple of ``trivial" exceptions: in the case of periodic boundary conditions there are no boundary blocks, so the map is defined as above, by neglecting the case (2c). Moreover, in the case of $+$ or $-$ boundary conditions, also the good boundary blocks have well defined boundary conditions, so also in this case (2c) should be neglected; e.g., if $B_i=[0,\ell_i]$ is a good boundary block adjacent to the left boundary of $[0,L]$ with $\o$ boundary conditions at its right end, and if the system has $+$ b.c.s imposed on the complement of $[0,L]$, then $B_i':=[-\frac12\log^2\ell_i,0]\cup B_i$ has $(+,\o)$ boundary conditions: therefore, we can define the replacement procedure in $B_i'$ as in items (2a)-(2b) above. 

%%%%%%%%%%%%%%%%%%%%%%%%%%%%%%%%%%%%%%%%%%%%%%%%%%%
\section{Proof of the main results.}
\label{sec:4}
%%%%%%%%%%%%%%%%%%%%%%%%%%%%%%%%%%%%%%%%%%%%%%%%%%%

In this section we prove Theorems \ref{thm00} and \ref{thm01} and Corollary \ref{cor1}, by using Proposition \ref{prop:0} and the explicit form of the coarse graining map $\phi\to\s_\phi$ described in the previous section. From now on, $C,C',\ldots$ and $c,c',\ldots$ indicate universal positive constants (to be thought of as ``big" and ``small", respectively), whose specific values may change from line to line.

Let us start with Theorem \ref{thm00}, whose proof is a simple corollary of Proposition \ref{prop:0}. In fact, using that that $e(h)-e(h^*)\ge 0$ and that $\widetilde F(x)\ge 0$, Eqs.(\ref{sp}) and (\ref{c000}) imply that $\mc E^{(\g)}_{[0,L]}[\phi]\ge Le(h^*)+O(\g^{1-\d})L$. This lower bound can be combined with the upper bound $\inf_\phi\mc E^{(\g)}_{[0,L]}[\phi]\le \mc E^{(\g)}_{[0,L]}[\phi^*]$, where, if $q_z(x)$ is the instanton defined after Eq.(\ref{2.inst}) and $z^*(x):=h^*\big(\lfloor x/h^*\rfloor+\frac12\big)$, then $\phi^*(x):=(-1)^{\lfloor x/h^*\rfloor}q_{z^*(x)}(x)$. An explicit computation shows that, for any $\e>0$, $\mc E^{(\g)}_{[0,L]}[\phi^*]\le Le(h^*)(1+C_\e\g^{1-\e})+C_b$ for two suitable constants $C_\e>0$ and $C_b$, where $C_b$ is a bound on the effect of the open boundary conditions. Combining the upper and lower bounds, we get Eq.(\ref{1.0}), as desired.

We now need to prove Theorem \ref{thm01}. We pick $\d_0$ and $\e_0$ so that $0<\d_0<\e_0<\frac13$. We also choose $\d$ and $\d_1$ in such a way that $0<\d<\min\{\frac13-\e_0,\d_0\}$ and $\d_0<\d_1<\frac23$. Here $\d$ indicates the coarse graining scale associated with the map $\phi\to\s_\phi$, which appears in the statement and proof of Proposition \ref{prop:0}.

If we combine the assumption Eq.(\ref{1.1}) (together with the known bound on $E_0^{(\g)}(L)$, following from Theorem \ref{thm00}) with the statement of Proposition \ref{prop:0}, we get,
\be
\label{4.1}
\begin{split}
L & e(h^*) +O(\g^{1-\d})L+\g^{\frac23+\e_0}L \; \ge \; \mc E^{(\g)}_{[0,L]}[\phi] \; \ge\;  L e(h^*)\\ & + \frac12\sum_{j=1}^{N_L^{\s_\phi}}h_j(e(h_j)-e(h^*))+ \frac{F(0)}{4 m_\b^2}\int_0^L\!\rmd x\,(|\s_\phi(x)|-m_\b)^2+O(\g^{1-\d})L\;,
\end{split}
\ee
where $h_j$ are the lengths of the maximal intervals on which $\s_\phi$ has constant sign. Note that, due to the assumption $0<\d<\frac13-\e_0$, the corrections $LO(\g^{1-\d})$ and $LO(\g^{4/3})$ are negligible with respect to the error term $L\g^{\frac23+\e_0}$. Therefore,
\be 
\label{4.2}
\frac12\sum_{j=1}^{N_L^{\s_\phi}}h_j(e(h_j)-e(h^*))+\frac{F(0)}{4 m_\b^2}\int_0^L\!\rmd x\,(|\s_\phi(x)|-m_\b)^2\le 2L\g^{\frac23+\e_0}.
\ee
Recall that $\s_\phi$ is associated with a partition $\BBB_\phi=\{B_i\}$ of $[0,L]$ such that all the blocks $B_i$ have lengths $\ell_i=|B_i|$ comparable with $\ell_+=\a_L(\d)\g^{-\d}$ (i.e., $\frac12\ell_+\le \ell_i\le \frac52\ell_+$); moreover, $\s_\phi$ is associated with a sequence of intervals $H_j$ on which $\s_\phi$ has constant sign. Given $\s_\phi$, we define a good set $\lis G_\phi$ as being the union of the intervals $H_j$  that are ``sufficiently long"; i.e., $\lis G_\phi=\cup_{j\colon h_j\ge \g^{-\d_1}} H_j$. It is easy to check that $\lis G_\phi^c:= [0,L]\setminus \lis G_\phi$ is just a small fraction of $[0,L]$. In fact, using the definition of $e(h)$, Eq.(\ref{eh}), we see that there exist a constant $c>0$ such that $e(h)-e(h^*)\ge \t/(2h)$ for all $h\le ch^*$ so that, using Eq.(\ref{4.2}),
\be 
\label{4.3}
\frac12\sum_{j\colon h_j\le \g^{-\d_1}}h_j\cdot\frac{\t}{2h_j}\le 2L\g^{\frac23+\e_0}\quad\Rightarrow\quad M_<^\mathrm{wrong}:=\sum_{j\colon h_j\le \g^{-\d_1}}1\le \frac{8}{\t}L\g^{\frac23+\e_0}
\ee
and, therefore, $|\lis G_\phi^c|=\sum_{j\colon h_j\le \g^{-\d_1}}h_j\le \frac{8}{\t}L\g^{\frac23+\e_0-\d_1}$. Let us now super-impose a coarser 
regular partition $\mc P_{\d_0}$ to the existing partition $\BBB_\phi$ (remember that $\d<\d_0$). Given $\phi$ and $\mc P_{\d_0}$, we let $W_\phi$ be the union of the elements of $\mc P_{\d_0}$ that have non zero intersection with $\lis G_\phi^c$. Of course, $|W_\phi|\le 2\g^{-\d_1}M_<^\mathrm{wrong}\le  \frac{16}{\t}L\g^{\frac23+\e_0-\d_1}$. Its complement, $W_\phi^c:=[0,L]\setminus W_\phi$, consists of a disjoint union of intervals $\L_k$, separated among each other by a distance larger than $\a_L(\d_0)\g^{-\d_0}$. The good set $G_\phi$ appearing in the statement of Theorem \ref{thm01} is defined as $G_\phi:=\{\L_k\colon |\L_k|\ge \g^{-\frac23-\frac{\e_0}2}\}$. Note that its complement, $G_\phi^c=[0,L]\setminus G_\phi$, is just a small fraction of $[0,L]$. In fact, $|G_\phi^c|\le |W_\phi|+\g^{-\frac23-\frac{\e_0}2}M_<^\mathrm{wrong}\le \frac{16}{\t}L\g^{\frac{\e_0}{2}}$, as stated in Theorem \ref{thm01}. 

Now, let us consider the blocks $Q_i\in\mc P_{\d_0}$ that are contained in $G_\phi$. As explained in the lines preceding Theorem \ref{thm01}, they can be of type $+$, $-$ or $0$. We group the $Q_i$'s contained in $G_\phi$ into maximal connected sequences of blocks of constant type, either $+$ or $-$, to be called $I_j$, $j=1,\ldots, \mc N_L^\phi$. By construction, the $I_j$'s within a single connected component $\L_k$ of $G_\phi$ have alternating sign and are separated among each other by at most one block of type $0$. We further denote by ${\rm int}\,I_j$ 
the subset of $I_j$ obtained from $I_j$ by depriving it of its first and last block in the sequence it consists of. Note that for each $I_j$ in $G_\phi$ there is an interval $H_j$ associated with the profile $\s_\phi$ such that ${\rm int}\,I_j\subseteq H_j$. This inclusion actually defines a one to one correspondence between the $I_j$'s and the intervals $H_j$ such that $H_j\cap G_\phi\neq \emptyset$, which justifies the use of the same label $j$ to indicate both $I_j$ and its corresponding $H_j$. Moreover, $I_j$, ${\rm int}\,I_j$ and $H_j$ have all essentially the same length, namely $|I_j|=h_j+O(\g^{-\d_0})$ and, similarly, $|{\rm int}\,I_j|= h_j+O(\g^{-\d_0})$. Note that $h_j\ge \g^{-\d_1}$, so that $|I_j|=h_j(1+o(1))$, with $|o(1)|\le \const \g^{\d_1-\d_0}$, and similarly for ${\rm int}\, I_j$.

We now want to bound from below the l.h.s of Eq.(\ref{4.2}) in terms of the left hand-side of Eq.(\ref{1.2}). Let us start with the second term in the left hand-side of Eq.(\ref{4.2}). Recall that $\widetilde F(\s):= \frac{F(0)}{2m_\b^2}(|\s|-m_\b)^2$. Obviously, 
\[
\int_0^L\!\rmd x\, \widetilde F(\s_\phi(x))\ge \sum_{j=1}^{\mc N_L^\phi} \int_{{\rm int}\, I_j}\!\rmd x\,\widetilde F(\s_\phi(x))\;.
\]
By construction, $\s_\phi$ has constant sign on each block $Q_i\subseteq I_j$. Since the restrictions of $\widetilde F(t)$ to the intervals $[-1,0]$ or $[0,1]$ are convex, we get,
\be 
\int_0^L\!\rmd x\,\widetilde F(\s_\phi(x))\ge \sum_{j=1}^{\mc N_L^\phi} \int_{{\rm int}\, I_j}\!\rmd x\,\widetilde F(\psi_\phi^{(\d_0)}(x))\;,
\ee
as desired. Consider now the first term in the left hand-side of Eq.(\ref{4.2}). We bound it as follows,
\be 
\sum_{j=1}^{N_L^{\s_\phi}}h_j(e(h_j)-e(h^*))\ge \sum_{\substack{j\colon H_j\cap G_\phi\neq\emptyset\\ |h_j-h^*|\ge h^*\g^{\e_0}}}h_j(e(h_j)-e(h^*) \;.
\ee
By using the definition of $e(h)$, Eq.(\ref{eh}), one can check that there exist constants $c,c',C,C'>0$ such that 
\be 
\label{4.6}
e(h)-e(h^*)\ge c\cdot\begin{cases} 1/h & {\rm if} \ \  h\le c'h^*\;,\\ \g^2(h-h^*)^2& {\rm if} \ \ c'h^*\le h\le C'\g^{-1}\;, \\ 1&{\rm if}\ \ h\ge C'\g^{-1}\;,
\end{cases}
\ee
and
\be
\label{4.7}
\big|e'(h)\big|\le C\cdot\begin{cases} 1/h^2 & {\rm if} \ \  h\le c'h^*\;,\\ \g^2|h-h^*|& {\rm if} \ \ c'h^*\le h\le C'\g^{-1}\;, \\ \g^{-1}/h^2&{\rm if}\ \ h\ge C'\g^{-1}\;. \end{cases} 
\ee
Using the fact that  $|I_j|=h_j+O(\g^{-\d_0})=h_j(1+o(1))$, we find,
\[
\begin{split}
& \sum_{\substack{j\colon H_j\cap G_\phi\neq\emptyset\\ |h_j-h^*|\ge h^*\g^{\e_0}}}h_j(e(h_j)-e(h^*)) \\ & \qquad \ge  \sum_{\substack{j\colon H_j\cap G_\phi\neq\emptyset \\  |h_j-h^*|\ge h^*\g^{\e_0}}}|I_j|(e(|I_j|)-e(h^*))\cdot \big(1+o(1)\big) \Big(1+\frac{e(h_j)-e(|I_j|)}{e(|I_j|)-e(h^*)}\Big)\;.
\end{split}
\] 
The error term $\frac{e(h_j)-e(|I_j|)}{e(|I_j|)-e(h^*)}$ can be estimated by making use of Eqs.(\ref{4.6})-(\ref{4.7}) as well as of the conditions $\big||I_j|-h_j\big|\le  2\g^{-\d_0}$ and $|h_j-h^*|\ge h^*\g^{\e_0}$. More precisely, if $h_j\le c' h^*$, then 
\[
\Big|\frac{e(h_j)-e(|I_j|)}{e(|I_j|)-e(h^*)}\Big|\le \Big|\int_{|I_j|}^{h_j}\!\rmd\x\, \big|e'(\x)\big|\frac1{e(|I_j|)-e(h^*)}\le \const \frac{\g^{-\d_0}}{h_j}\le \const\g^{\d_1-\d_0}\;.
\]
Similarly, if $c' h^*\le h_j\le C'\g^{-1}$ and $|h_j-h^*|\ge h^*\g^{\e_0}$, then 
\[ 
\Big|\frac{e(h_j)-e(|I_j|)}{e(|I_j|)-e(h^*)}\Big|\le \const\g^{-\d_0}\frac{\g^2|h_j-h^*|}{\g^2(h_j-h^*)^2} \le \const\g^{\frac23-\d_0-\e_0}\;,
\]
while, if $h_j\ge C'\g^{-1}$, 
\[ 
\Big|\frac{e(h_j)-e(|I_j|)}{e(|I_j|)-e(h^*)}\Big|\le \const\g^{-\d_0}\frac{1}{\g h_j^2}\le \const\g^{1-\d_0}\;.
\]
In conclusion,
\be 
\label{4.12}
\sum_{j=1}^{N_L^{\s_\phi}}h_j(e(h_j)-e(h^*))\ge \frac12  \sum_{\substack{j\colon H_j\cap G_\phi\neq\emptyset\\ |h_j-h^*|\ge h^*\g^{\e_0}}}|I_j|(e(|I_j|)-e(h^*))\;.
\ee
We now want to add back the sum over the intervals such that $|h_j-h^*|< h^*\g^{\e_0}$ in the right hand-side of Eq.(\ref{4.12}). To this purpose, note that 
\[
\sum_{\substack{j\colon H_j\cap G_\phi\neq\emptyset\\ |h_j-h^*|< h^*\g^{\e_0}}}|I_j|(e(|I_j|)-e(h^*))\le 
\const L\g^2(h^*)^2\g^{2\e_0}\le \const L\g^{\frac23+2\e_0}\;,
\]
so that, finally,
\[ 
\sum_{j=1}^{N_L^{\s_\phi}}h_j(e(h_j)-e(h^*))\ge \frac12  \sum_{j=1}^{\mc N_L^\phi}|I_j|(e(|I_j|)-e(h^*))-O(L\g^{\frac23+2\e_0})\;.
\]
Putting all together gives Eq.(\ref{1.2}). \qed

\medskip
Let us conclude this section by proving Corollary \ref{cor1}. We start with the estimate on $X_1$. By Eq.(\ref{1.2}) we have,
\be 
\label{4.15}
\sum_{j=1}^{\mc N_L^\phi} \int_{{\rm int}\,I_j}\!\rmd x\, (|\psi_{\phi}^{(\d_0)}(x)|-m_\b)^2\le \const L\g^{\frac23+\e_0}\;.
\ee
The left hand-side can be bounded from below by
\[
\sum_{j=1}^{\mc N_L^\phi}\int_{{\rm int}\,I_j\cap X_1}\!\rmd x\, (|\psi_{\phi}^{(\d_0)}(x)|-m_\b)^2\ge \g^{2\e}|X_1|\;,
\]
that, if combined with Eq.(\ref{4.15}), gives the first of Eq.(\ref{1.3}). Let us now turn to the estimate on $X_2$. By Eq.(\ref{1.2}) we have,
\be 
\label{4.17}
\sum_{j\colon I_j\in X_2}|I_j|(e(|I_j|)-e(h^*))\le 10 L\g^{\frac23+\e_0}\;.
\ee
Using Eqs.(\ref{4.6})-(\ref{4.7}), the left hand-side can be bounded from below by
\[
\const\Big[\sum_{|I_j|\le c'h^*}|I_j|\frac1{|I_j|}+\hskip-.1truecm \sum_{\substack{c'h^*\le |I_j|\le C'\g^{-1}\\
||I_j|-h^*|\ge h^*\g^{\e'}}} \hskip-.1truecm |I_j|\g^2(|I_j|-h^*)^2+\sum_{|I_j|\ge C'\g^{-1}}|I_j|\Big]\;,
\]
which is larger than $\g^{\frac23+2\e'}|X_2|$. Combining this with Eq.(\ref{4.17}) gives the second of Eq.(\ref{1.3}).
\qed

%%%%%%%%%%%%%%%%%%%%%%%%%%%%%%%%%%%%%%%%%%%%%%%%%%%
\section{Proof of Proposition \ref{prop:0}}
\label{sec:5}
%%%%%%%%%%%%%%%%%%%%%%%%%%%%%%%%%%%%%%%%%%%%%%%%%%%

In this section we prove Proposition \ref{prop:0}. We proceed in two main steps: we first prove Eq.(\ref{sp}), which allows us to replace the original functional $\mc E^{(\g)}_{[0,L]}[\phi]$ by the effective functional $\widetilde{\mc E}^{(\g)}_{[0,L]}[\s_\phi]$, provided that $\s_\phi$ is chosen in the way described in Subsection \ref{subsec3.2}; next, we study the effective functional by reflection positivity methods and we prove Eq.(\ref{c000}). A prominent role in the proof of the first step is played by the following two propositions, whose proof is deferred to Appendix \ref{sec:C}. Hereafter, we shall denote by $\{b_j\}$ a partition of $\mathbb R$ into a sequence of small blocks $b_j$ of size $\{2^{-n}, n\in\mathbb{N}\}$ (the choice that $\ell_-=2^{-n}$ for some $n$ guarantees that $b_j$ is divisible by the range of $J$, which is 1).
\begin{proposition}
\label{prop:1}
There are constants $\zeta_0, \kappa_0, \alpha,C_0$, all positive, such that the following holds. Given $\z_1<\z_0$, $\ell_-<\k_0\z_1$, and an interval $B=[a,a+\ell]$, of size $\ell$ large enough, suppose that $\phi\in L^\infty(\mathbb{R};[-1,1])$ satisfies, for some $\o_\pm\in\{\pm1\}$, 
\[
\begin{split}
& |\media{\phi}_{b_j}-\o_-m_\b|\le \zeta_1 \quad \forall\,b_j\subset [a,a+2\log^2\ell]\;, \\ & |\media{\phi}_{b_j}-\o_+m_\b|\le \zeta_1 \quad \forall\,b_j\subset [a+\ell-2\log^2\ell,a+\ell]\;. \\ 
\end{split}
\]
Then there exists $\tilde\phi\in L^\infty(\mathbb{R};[-1,1])$ such that\\
\begin{equation}
\label{bp1}
\mc E_B^{(0)}[\tilde\phi] \le \mc E_B^{(0)}[\phi]+ C_0\, \mathrm{e}^{-2\alpha\,\log^2\ell}\;, 
\end{equation}
\begin{equation}
\label{bp2}
|\media{\tilde\phi}_B-\media{\phi}_B| \le 8\zeta_1\frac{\log^2\ell}\ell\;,
\end{equation}
\begin{equation}
\label{bp3}
\begin{split}
&\tilde\phi(x) = \omega_-m_\b \quad\forall\,x\in (-\infty,a+\log^2\ell]\;, \\ & \tilde\phi(x) = \omega_+m_\b \quad\forall\,x\in [a+\ell-\log^2\ell,\infty)\;.
\end{split}
\end{equation}
\end{proposition}
\begin{proposition}
\label{prop:2} 
There are constants $\z_0,\k_0, C_0$, all positive, such that the following holds. Let $B$ be an interval of size $\ell$ and let $\mc T$ be the one dimensional torus of size $\ell$. Given $\z_1<\z_0$ and $\ell_-<\k_0\z_1$, let $\phi_\o\in L^\io(\mathbb R;[-1,1])$ be such that $|\media{\phi_\o}_{b_j}-\o m_\b|\le \z_1$, with $\o\in\{\pm\}$. Let us indicate by $\mc E^{(0)}_B[\phi]$ the functional (\ref{00.2}) on $B$ with $\g=0$ and open boundary conditions; similarly, let us indicated by $\mc E^{(0);\phi_\o}_{B}[\phi]$ the functional on $B$ with $\g=0$ and $\phi_\o$ boundary conditions and by $\mc E_{\mc T}^{(0);\mathrm{per}}[\phi]$ the functional on $\mc T$ with $\g=0$ and periodic boundary conditions on $\mc T$. Then, for any $\ell$ large enough the following holds.
\\ \noindent
(I) If $|m|\ge m_\b$, the unique minimizer for $\mc E_{B}^{(0)}[\phi]$ with average $\media{\phi}_B=m$ is the uniform profile $\phi(x)=m$.
\\ \noindent
(II) If  
\[
m_\b - \frac{\log^3\ell}\ell \le |m| \le m_\b
\]
the unique minimizer for $\mc E_{B}^{(0);\phi_{m/|m|}}[\phi]$ with average $\media{\phi}_B=m$ is the uniform profile $\phi(x)=m$.
\\ \noindent
(III) The energy $\mc E_{\mc T}^{(0);\mathrm{per}}[\phi]$ of any profile $\phi$ with average $\media{\phi}_B=m$ is bounded from below as
\be 
\label{i4} 
\mc E_{\mc T}^{(0);\mathrm{per}}[\phi] \ge \min\{\ell F(m), 2\t- C_0\rme^{-2\a\log^2\ell}\}\;.
\ee
\end{proposition}
We are now finally ready to describe the two main steps entering the proof of Proposition \ref{prop:0}.

\subsection{The replacement procedure: proof of Eq.(\ref{sp})}
First of all, observe that $\s_\phi$ has the property that $\media{\phi}_{B_i}=\media{\s_\phi}_{B_i}$ for all the elements $B_i\in\mc B_\phi$. Due to this fact, the difference between the contribution to $\mc E^{(\g)}_{[0,L]}[\phi]$ coming from the long range potential (we shall call it the ``dipole energy" and denote it by $\mc V^{(\g)}_{[0,L]}[\phi]$) and $\mc V^{(\g)}_{[0,L]}[\s_\phi]$ is small: in fact, it is easy to check that given $\mc B_\phi$ and two arbitrary functions $\phi_1$ and $\phi_2$ such that $\phi_1\big|_{B_i^c}=\phi_2\big|_{B_i^c}$ and $\media{\phi_1}_{B_i}=\media{\phi_2}_{B_i}$ for some $B_i\in\mc B_\phi$, then $\mc V^{(\g)}_{[0,L]}[\phi_1]-\mc V^{(\g)}_{[0,L]}[\phi_2]=|B_i|O(\g |B_i|)$; therefore, replacing $\phi\to\s_\phi$ in one block $B_i$ at a time, for all $B_i\in\mc B_\phi$, gives,
\be 
\big|\mc V^{(\g)}_{[0,L]}[\phi]-\mc V^{(\g)}_{[0,L]}[\s_\phi]\big|\le \const L\g^{1-\d}\;,
\ee
so that 
\be 
\mc E^{(\g)}_{[0,L]}[\phi]\ge \mc E^{(0)}_{[0,L]}[\phi]+\mc V^{(\g)}_{[0,L]}[\s_\phi]+O(L\g^{1-\d})\;.
\ee
Moreover, by neglecting the short range interaction between contiguous segments $V_k$ (for a definition of $V_k$, see the discussion after Lemma \ref{lem:1}), we get,
\be  
\mc E^{(\g)}_{[0,L]}[\phi]\ge \sum_{k}\mc E^{(0)}_{V_k}[\phi]+\mc V^{(\g)}_{[0,L]}[\s_\phi]+O(L\g^{1-\d})\;.
\ee

We are left with proving that, for all the segments $V_k$, $\mc E^{(0)}_{V_k}[\phi]$ is bounded from below by $\widetilde{\mc E}^{(0)}_{V_k}[\s_\phi]$, up to an error term smaller than $|V_k|O(\g^{1-\d})$. There are two cases: either $V_k$ consists of several bad blocks, or $V_k$ consists of exactly one good block.

\subsubsection{Case 1: $V_k$ has more than one block in its interior.}
Remember that by construction all the blocks in $V_k$ have internal energy larger than $2\t$ and that the replacement procedure $\phi\to\s_\phi$ in $B_i\subset V_k$ (defined in item (1) of Subsection \ref{subsec3.2}) is the following: if $|m_i|\ge m_\b-\z$, then $\s_\phi=m_i$ on $B_i=[a,a+\ell_i]$; if $|m_i|<m_\b-\z$, then $\s_\phi=m_\b$ on $[a,a+\x]$ and $=-m_\b$ on $[a+\x,a+\ell_i]$, with $\x$ fixed in such a way that $\media{\s_\phi}_{B_i}=m_i$.  Let $M^+_k$ (resp.\ $M^-_k$, resp. $M^0_k$) be the number of blocks $B_i\subset V_k$ such that $m_i\ge m_\b-\z$ (resp.\ $m_i\le -m_\b+\z$, resp. $|m_i|<m_\b-\z$). Obviously, 
\be 
\label{5.10}
\mc E^{(0)}_{V_k}[\phi]\ge \sum_{\substack{B_i\subset V_k\colon \\|m_i|\ge m_\b-\z}}\mc E^{(0)}_{B_i}[\phi]+2\t M^0_k\;,
\ee
simply because every block $B_i$ carries an internal energy larger than $2\t$. Moreover, if $|m_i|\ge m_\b-\z$ and $\mc E^{(0)}_{B_i}[\phi]>2\t$, we have that $\mc E_{B_i}^{(0)}[\phi]\ge |B_i|\widetilde F(m_i)+\t$, with $\widetilde F(m)=\frac{F(0)}{2m_\b^2}(|m|-m_\b)^2$. In fact, if $|m_i|\ge m_\b$, by item (I) of Proposition \ref{prop:2}, 
\be
\label{5.11}
\mc E_{B_i}^{(0)}[\phi] \ge \frac 12 \mc E_{B_i}^{(0)}[m_i]+\tau =\frac12|B_i|F(m_i)+\t\ge  |B_i| \widetilde F(m_i) + \tau\;,
\ee
where in the last inequality we used the fact that $\frac12F(t)\ge \widetilde F(t)$, see Remark \ref{rem:p2} and Appendix \ref{sec:A}. If, on the other hand, $m_\b-\z\le |m_i|\le m_\b$, then $\widetilde F(m_i)\le \frac{F(0)}{2 m_\b^2}c_0^2\g^{2\d}\log^4\g$, so that $|B_i|\widetilde F(m_i)\ll\t$ and, therefore, 
\be 
\label{5.12}
E_{B_i}^{(0)}[\phi] \ge |B_i|\widetilde F(m_i)+\t\;.
\ee
Plugging Eqs.(\ref{5.11})-(\ref{5.12}) into Eq.(\ref{5.10}) gives,
\[
\begin{split}
\mc E^{(0)}_{V_k}[\phi]&\ge \sum_{\substack{B_i\subset V_k\colon \\|m_i|\ge m_\b-\z}}|B_i|\widetilde F(m_i)+\t M^+_k+\t M^-_k+2\t M^0_k \\
&=\int_{V_k}\!\rmd x\,\widetilde F(\s_\phi)+\t M^+_k+\t M^-_k+2\t M^0_k\;.
\end{split}
\]
Now note that $M^+_k+M^-_k+2M^0_k$ is larger than the total number of jumps of $\s_\phi$ in 
$V_k$ (let us call it $M^\phi_k$) and, therefore, we get the desired estimate,
\[
\mc E^{(0)}_{V_k}[\phi]\ge \int_{V_k}\!\rmd x\, \widetilde F(\s_\phi)+\t M^\phi_k=\widetilde{\mc E}^{(0)}_{V_k}[\s_\phi]\;.
\]
\subsubsection{Case 2: $V_k$ has a single block in its interior.} 
In this case $V_k$ consists of a single good block, which we call $B$ and denote its length by $\ell$ (and we write $B=[a,a+\ell]$ and $m=\media{\phi}_B$). We need to prove that $\mc E^{(0)}_B[\phi]\ge \widetilde{\mc E}^{(0)}_B[\s_\phi]+\ell O(\g\ell)$. Remember that the replacement procedure $\phi\to\s_\phi$ in $B$ was defined in items (2a), (2b) and (2c) of Subsection \ref{subsec3.2}) and depends on the specific boundary conditions assigned to $B$. Let us discuss these three cases separately.
\\
a) Suppose that $B$ has boundary conditions $(\o,-\o)$. In this case, if $|m|\le m_\b-\z$, then $\s_\phi=\o m_\b$ on $[a,a+\x]$ and $=-\o m_\b$ on $[a+\x,a+\ell_i]$, with $\x$ fixed in such a way that $\media{\s_\phi}_{B}=m$; if $|m|>m_\b-\zeta$, then 
\[
\sigma_\phi(x) = \begin{cases} 
\o m_\b & {\rm if}\quad x\in [a, a+\frac12\log^2\ell] \;,\\  
m\ell(\ell-\log^2\ell)^{-1} & {\rm if}\quad x\in(a+\frac12\log^2\ell,a+\ell-\frac12\log^2\ell)\;,\\ 
-\o m_\b & {\rm if}\quad x\in[a+\ell-\frac12\log^2\ell,a+\ell]\;.\end{cases}
\]
Let us start by considering the case that $|m|\le m_\b-\z$. By Proposition \ref{prop:1} with $\z_1$ small enough (but independent of $\g$), there exists $\tilde \phi:\mathbb R\to[-1,1]$ such that $\tilde\phi(x)=\o m_\b$ for all $x\le a+\log^2\ell$, $\tilde\phi(x)=-\o m_\b$ for all $x\ge a+\ell-\log^2\ell$, and
\be 
\mc E^{(0)}_B[\phi]\ge \mc E^{(0)}_{\mathbb R}[\tilde \phi]-C_0\rme^{-2\a\log^2\ell}\ge 
\t-C_0\rme^{-2\a\log^2\ell}\;,
\ee
where in the second inequality we used Eq.(\ref{2.1}). Noting that $\t=\widetilde{\mc E}^{(0)}_B[\s_\phi]$, we have the desired estimate. Let us now consider the case $|m|\ge m_\b-\z$ and let us assume without loss of generality that $m>m_\b-\zeta$. Setting 
\[
\mathcal{I}_+ = \{x\in B\colon \phi(x)\ge 0\}, \quad \mathcal{I}_-=\{x\in B\ \colon \phi(x)< 0\}\;, \quad m_\pm = \pm\media{\phi}_{\mathcal{I}_\pm}
\]
and defining $m_* = \ell^{-1}|\mathcal{I}_+| m_++\ell^{-1}|\mathcal{I}_-|m_-$, since $\widetilde F$ is an even function, convex on $[0,1]$, and using the fact that $|\mathcal{I}_+| +|\mathcal{I}_-| = \ell$, we get
\be
\label{p1b}
\begin{split}
\int_B\!\rmd x\, F(\phi(x)) & \ge \int_B\!\rmd x\, 2\widetilde F(\phi(x)) \ge 2|\mathcal{I}_+|\widetilde F(m_+) + 2 |\mathcal{I}_-| \widetilde F(m_-) \\ & \ge\; 2\ell \widetilde F(m_*) = 2\ell \widetilde F(m +2\ell^{-1}|\mathcal{I}_-|m_-)\;.
\end{split}
\ee
Recalling that the boundary condition is negative on one side of $B$, Lemma \ref{lem:1} implies that $|\mathcal{I}_-|m_- \ge c \gamma^{-(\d -2\rho)}$, whence 
\be 
\label{mlr}
m +2\ell^{-1}|\mathcal{I}_-|m_- \ge m + 2c' \gamma^{2\rho} \ge m_\b + c' \gamma^{2\rho}\;,
\ee
where in the last inequality we used the assumption $m\ge m_\b -\zeta$. By the definition of $\widetilde F$, we conclude that
\be
\int_B\!\rmd x\, F(\phi(x)) \ge 2\ell\widetilde F(m_*)\ge c'' \ell \gamma^{4\rho}\; .
\label{p2}\ee
Choosing $\rho<\d/4$, the right hand-side is much bigger than $2\t$ for any $\gamma$ small enough. Note also that, using the fact that $\ell^{-1}|\mc I_-|m_-\ge \const\g^{2\r}$, we get $m\ell/(\ell-\log^2\ell)\le (m_*-c'''\g^{2\r})(1+C\g^\d\log^2\g)\le m_*<1$; moreover, 
using that $m\ge m_\b-\z$, we have $m\ell/(\ell-\log^2\ell)\ge (m_\b-c_0\g^\d\log^2\g)(1+C\g^\d\log^2\g)$, which is larger than $m_\b$ for $c_0$ small enough. Therefore, $m_\b<\frac{m\ell}{\ell-\log^2\ell}<m_*$, so that $\widetilde F(\frac{m\ell}{\ell-\log^2\ell})<\widetilde F(m_*)$. Putting this
together with Eq.(\ref{p2}) gives:
\be \int_B\!\rmd x\, F(\phi(x)) \ge\ell\widetilde F(m_*)+\t\ge \int_B\!\rmd x\, \widetilde F(\s_\phi(x))+\t =
\widetilde{\mc E}_B^{(0)}[\s_\phi]\;,\ee
as desired.
\\
b) Suppose that $B$ is not a boundary block and that its boundary conditions are $(\o,\o)$. Without loss of generality, we assume that $(\o,\o)=(-,-)$. We recall that $\s_\phi$ is defined as follows:
if $-1\le m\le - m_\b + \frac{11}{10}C_*\ell^{-1/2}$ then $\s_\phi(x)\equiv m$;
if $ - m_\b +\frac{11}{10} C_*\ell^{-1/2}<m<m_\b-\frac{11}{10}C_*\ell^{-1/2}$, then $\s_\phi(x)=-m_\b$ on $[a,a+\x]\cup[a+\ell-\x,a+\ell]$ and $=m_\b$ on $[a+\x,a+\ell-\x]$, with 
$\x$ fixed in such a way that $\media{\s_\phi}_{B}=m$;
if $m > m_\b - \frac{11}{10}C_*\ell^{-1/2}$, then
\be 
\label{15bis.m}
\sigma_\phi(x) = \begin{cases} \frac{m\ell + m_\b \log^2\ell}{\ell-\log^2\ell} & {\rm if}\quad x\in 
(a+\frac12\log^2\ell,a+\ell-\frac12\log^2\ell)\;,\\ 
- m_\b & {\rm if}\quad x\in [a,a+\frac12\log^2\ell]\cup[a+\ell-\frac12\log^2\ell,a+\ell]\;.\end{cases}
\ee
In order to show that $\mc E^{(0)}_B[\phi]$ is bounded from below by $\widetilde{\mc E}^{(0)}_B[\s_\phi]$ up to subdominant error terms, we proceed separately in three subcases.\\
(b.1)
If $-1\le m \le - m_\b + \ell^{-1}\log^3\ell$, by items (I) and (II) of Proposition \ref{prop:2}, we immediately get that $\mc E^{(0)}_B[\phi]\ge \mc E^{(0)}_B[m]\ge \ell\widetilde F(m)=\widetilde{\mc E}^{(0)}_B[\s_\phi]$.\\
(b.2) if $ - m_\b + \ell^{-1}\log^3\ell< m \le m_\beta - \frac{11}{10}C_*\ell^{-1/2}$ we apply Proposition \ref{prop:1} and denote by $\hat\phi$ the $\ell$-periodic function obtained by periodization of the profile $\tilde\phi(x)$ appearing in the statement of that proposition. Recalling \eqref{bp1}, \eqref{bp2}, and \eqref{bp3}, we have
\begin{equation}
\label{hat}\mc E_B^{(0)}[ \phi] \ge 
\mc E_\mathcal{T}^{(0);\mathrm{per}}[\hat\phi] - C_0\, \rme^{-2\alpha\,\log^2\ell}\;, \qquad |m-\hat m| \le  8\z_1\frac{\log^2\ell}{\ell}\;,
\end{equation}
where $\hat m = \media{\hat\phi}_\mathcal{T}$. 

Consider first the case that $-m_\b+\frac{\log^3\ell}{\ell}\le m\le -m_\b+\frac{11}{10}C_*\ell^{-1/2}$, in which case we want to prove that $\mc E^{(0)}_B[\phi]\ge \ell \widetilde F(m)$. By the second inequality in Eq.(\ref{hat}), $-m_\b+\frac12\frac{\log^3\ell}{\ell}\le \hat m\le -m_\b+\frac{6}{5}C_*\ell^{-1/2}$. We have three more sub cases:
\begin{itemize}
\item If $-m_\b+\frac12\frac{\log^3\ell}{\ell}\le \hat m\le -m_\b+c_*\ell^{-1/2}$ then by item (III) of Proposition \ref{prop:2} we have that $\mc E_B^{(0)}[ \phi] \ge \ell F(\hat m)- C_0\, \rme^{-2\alpha\,\log^2\ell}$, with 
\[
\ell F(\hat m)\ge\frac{\ell F(0)}{m^2_\b}(\hat m+m_\b)^2
\ge \frac1{10} \frac{\ell F(0)}{m^2_\b}\frac{\log^6\ell}{4\ell^2}+\frac9{10}\frac{\ell F(0)}{m_\b^2} \big[(\hat m-m)+(m+m_\b)\big]^2\;,
\] 
where in the last inequality we used that $\hat m+m_\b\ge \frac12\frac{\log^3\ell}{\ell}$; moreover, using that for all $\e\in(0,1)$ one has $(a+b)^2\ge (1-\e)a^2+(1-\e^{-1})b^2$, we get (fixing $\e=\frac1{10}$), 
\[
\ell F(\hat m)\ge  \frac1{10} \frac{\ell F(0)}{m^2_\b}\frac{\log^6\ell}{4\ell^2}+\frac{81}{100}\frac{\ell F(0)}{m_\b^2}(m+m_\b)^2-c\z_1^2\frac{\log^4\ell}{\ell^2}\;,
\]
where we also used the second of Eq.(\ref{hat}). Putting all together,
\[
\mc E_B^{(0)}[ \phi] \ge  \frac1{10} \frac{\ell F(0)}{m^2_\b}\frac{\log^6\ell}{4\ell^2}+\frac{81}{100}\frac{\ell F(0)}{m_\b^2}(m+m_\b)^2-c\z_1^2\frac{\log^4\ell}{\ell^2}-C_0\rme^{-2\a\log^2\ell}\ge 
\ell\widetilde F(m)\;,
\]
as desired.
\item  If $-m_\b+c_*\ell^{-1/2}\le \hat m\le -m_\b+C_*\ell^{-1/2}$ then by item (III) of Proposition \ref{prop:2}, $\mc E_B^{(0)}[ \phi] \ge \min\{2\t, \ell F(\hat m)\}+O(\rme^{-2\a\log^2\ell})$. Using the fact that $F''(m_\b)\ge \frac{2F(0)}{m_\b^2}$ (see Appendix \ref{sec:A}) and $(\hat m+m_\b)^2\le C_*^2\ell^{-1}=\frac{ 5\t }{ F''(m_\b)}\ell^{-1}$, we get $2\t\ge\frac{ 4\t F(0)}{m_\b^2 F''(m_\b)}\ge \frac45\frac{\ell F(0)}{m_\b^2} (\hat m+m_\b)^2$. Combining this with $\ell F(\hat m)\ge \frac{\ell F(0)}{m_\b^2}(\hat m+m_\b)^2$, we find
\[
\mc E_B^{(0)}[ \phi] \ge \frac45\frac{\ell F(0)}{m_\b^2} (\hat m+m_\b)^2-C\rme^{-2\a\log^2\ell}\;.
\]
Proceeding as in the previous item, we can further bound this by
\[
\begin{split}
\mc E_B^{(0)}[ \phi] & \ge \ell F(\hat m)\ge   \frac2{25} \frac{c_*^2 F(0)}{m^2_\b}+\frac{81}{125}\frac{\ell F(0)}{m_\b^2}(m+m_\b)^2-c\z_1^2\frac{\log^4\ell}{\ell^2}-C\rme^{-2\a\log^2\ell} \\ & \ge \ell \widetilde F(m)\;,
\end{split}
\]
as desired.
\item If $-m_\b+C_*\ell^{-1/2}\le \hat m\le -m_\b+\frac{6}{5}C_*\ell^{-1/2}$, then by item (III) of Proposition \ref{prop:2} $\mc E^{(0)}_B[\phi]\ge 2\t-C_0' \rme^{-2\a\log^2\ell}$. On the other hand $\ell\widetilde F(m)=\frac{\ell F(0)}{2m_\b^2}(m+m_\b)^2\le \big(\frac{11}{10}\big)^2\frac{5\t F(0)}{2m_\b^2 F''(m_\b)}\le \frac{121}{80}\t$, so that 
\[
\mc E^{(0)}_B[\phi]\ge \frac{160}{121}\ell\widetilde F(m)-C_0' \rme^{-2\a\log^2\ell}>\ell\widetilde F(m)\;,
\] 
as desired.
\end{itemize}
Finally, consider the case that $-m_\b+\frac{11}{10}C_*\ell^{-1/2}\le m\le m_\b-\frac{11}{10}C_*\ell^{-1/2}$, in which case we want to prove
that $\mc E^{(0)}_B[\phi]$ is bounded from below by $2\t$ up to subdominant terms. By the second inequality in Eq.(\ref{hat}), $-m_\b+C_*\ell^{-1/2}\le m\le m_\b-C_*\ell^{-1/2}$. Therefore, we can apply item (III) of Proposition \ref{prop:2} to conclude that $\mc E^{(0)}_B[\phi]\ge 2\t-C_0'\rme^{-2\a\log^2\ell}$, as desired.
\\
(b.3) If $m > m_\b - \frac{11}{10}C_*\ell^{-1/2}$, then we want to show that $\mc E^{(0)}[\phi]\ge 2\t+(\ell-\log^2\ell)\widetilde F(\frac{m\ell+m_\b\log^2\ell}{\ell-\log^2\ell})$. By Eqs.(\ref{p1b})--(\ref{p2}), with $\r<\d/4$, we have that $\int_B\!\rmd x\,F(\phi(x))\ge 2\ell \widetilde F(m_*)\ge c\ell\g^{4\r}\ge 4\t$ and $m_*\ge m+c\g^{2\r}\ge m_\b+c'\g^{2\r}$. Moreover, $\frac{m\ell+m_\b\log^2\ell}{\ell-\log^2\ell}\le m+C\g^\d\log^2\g<m_*$, so that 
\[
\int_B\!\rmd x\, F(\phi(x))\ge 2\t+\ell\widetilde F(m_*)\ge 2\t+\ell\widetilde F(m)\;,
\]
as desired.
\\
c) If $B$ is a good boundary block, we proceed in a way very similar to item (b). Let us assume without loss of generality that $B=[0,\ell]$ is the boundary block at the left of the interval $[0,L]$ and that its right boundary condition is $\o$ (while by construction its left boundary condition is open). Let us also define $B_1=[-\ell,0]$ and $B_2=[-\ell,\ell]$. Given $\phi:B\to[-1,1]$ with $\o$ boundary conditions on the right end of $B$, we define $\phi_2:B_2\to[-1,1]$ to be $\phi_2(x)=\phi(|x|)$. Of course,
\[ 
\mc E^{(0)}_B[\phi]=\frac12\mc E^{(0)}_{B_2}[\phi_2]-\frac12\int_0^1\!\rmd x \int_0^1\! \rmd y\, J(x+y)[\phi(y)-\phi(x)]^2\ge \frac12\mc E^{(0)}_{B_2}[\phi_2]-C\;.
\]
Now $\mc E^{(0)}_{B_2}[\phi_2]$ can be bounded from below exactly as discussed in item (b), and we are left with the desired estimate up to an error term of order 1 (the constant $-C$ appearing in the right hand-side of the last equation). However, there are at most two such contributions from the whole interval $[0,L]$. Therefore these two error terms of order 1 can be reabsorbed into the overall error $O(\g^{1-\d})L$ provided $L\g\gg 1$, as assumed in the statement of Proposition \ref{prop:0}. This concludes the proof of Eq.(\ref{sp}).
\qed

\subsection{Reflection positivity: proof of Eq.(\ref{c000})} 
In order to bound $\widetilde {\mc E}^{(\g)}_{[0,L]}[\s]$ from below, we use the ideas and methods of \cite{GLL1,GLL07,GLL2,GLL3,GMu,GLL11}. The first step consists in using the {\it chessboard estimate with open boundary conditions} (see e.g. \cite[Appendix A]{GLL07}, \cite[Section 3]{GLL2} or \cite[Appendix A]{GMu}). Roughly speaking, we repeatedly reflect $\s$ at the jump points of $\s$ (where $\s$ goes from a positive to a negative value, or viceversa), after which 
\be 
\label{3.1}
\widetilde{\mc E}^{(\g)}_{[0,L]}[\s]\ge \sum_{i=1}^{N_L^\s}h_i \widetilde e_{h_i}[\widetilde\s_i]\;,
\ee
where: $h_i$ are the length of the intervals $H_i$ on which $\s$ has constant sign; $\widetilde\s_i$ is the periodic function on the real line consisting of blocks all of length $h_i$ and alternating sign, such that $\widetilde\s_i\big|_{H_i}=\s\big|_{H_i}$ (i.e., it is the ``anti periodic" extension of the restriction of $\s$ to the interval $H_i$); $\widetilde e_{h_i}[\widetilde\s_i]$ is the specific energy of such configuration computed by using the functional $\widetilde{\mc E}^{(\g)}_{\mathbb R}$. An explicit computation gives (assuming without loss of generality that $\s$ is positive on $[0,h]$),
\be 
\label{3.2}
\widetilde e_h[\s]=\frac1h\int_0^h\!\rmd x\, \widetilde F(\s(x)) +\frac{\t}h +\frac{1}{2 h}\int_0^h\! \rmd x \int_0^h\!\rmd y\, \s(x) \widetilde v_h(x,y)\s(y)\;,
\ee
where (see \cite[Eq.(3.24)]{GLL2}) 
\be  
\label{1.200}
\widetilde v_h(x,y) =\g \sum_{n\in\bb Z}\Big[v(\g(2nh+y-x))-v(\g(2nh+y+x))\Big]\;,
\ee
which can be equivalently rewritten as
\be
\label{1.20a}
\begin{split}
\widetilde v_h(x,y) & =\g\sum_{n\ge 0}\Big[v(\g(2nh+y-x))-v\big(\g(2(n+1)h-y-x)\big) \\ & \quad + v\big(\g(2(n+1)h-y+x)\big)-v(\g(2nh+y+x))\Big]\;.
\end{split}
\ee
Two useful remarks are the following:
\begin{enumerate}
\item since $v$ is convex, then each term in square brackets in the right-hand side of Eq.\eqref{1.20a} is positive: therefore, $\widetilde v_h(x,y)$ is pointwise positive;
\item the quadratic form defined by $\widetilde v_h$ is positive definite (simply because the potential $v(x-y)$, which $\widetilde v_h$ is constructed from, is positive definite), so that 
\be 
\label{3.5}
(f,g)_{\tilde v_h}:=\int_0^h\!\rmd x\int_0^h\!\rmd y\,f(x)\widetilde v_h(x,y)\, g(y)
\ee
defines a scalar product; in the following we shall denote by $\|\cdot\|_{\tilde v_h}$ the norm induced by this scalar product.
\end{enumerate}
Let us now distinguish two cases. If $h\ge C\g^{-2/3}$, with $C$ a suitable (sufficiently large) $O(1)$ constant, then using the pointwise positivity of $\widetilde v_h$ and the fact that $\s(x)\ge \lis m:=m_\b-\k\g^{\d/2}$ on $[0,h]$, we get:
\[ 
\widetilde e_h[\s]\ge \frac1h\int_0^h\!\rmd x\, \widetilde F(\s(x)) +\frac{\t}h +\frac{1}{2 h}(\lis m,\lis m)_{\widetilde v_h}\;.
\]
Using the explicit expression for $(\lis m,\lis m)_{\widetilde v_h}$ and Eq.(\ref{eh}), we can rewrite the latter estimate as
\[
\begin{split}
& \widetilde e_h[\s]\ge e(h^*)+\frac1h\int_0^h\!\rmd x\, \widetilde F(\s(x)) \\ & + \frac{\t}h-\frac{\t}{h^*} +\l\int\!\frac{\m (\rmd\a)}{\a}\Big[\lis m^2\Big(1-\frac{\tanh(\a\g h/2)}{\a\g h/2}\Big)- m_\b^2\Big(1-\frac{\tanh(\a\g h^*/2)}{\a\g h^*/2}\Big)\Big]\;.
\end{split}
\]
It is easy to check that the expression in the second line is bounded from below by $\frac12(e(h)-e(h^*))$. In fact, this is equivalent to the condition, 
\[
\begin{split}
& \l\Big(\lis m^2-\frac{m_\b^2}2\Big)\int\!\frac{\m (\rmd\a)}{\a} \Big(1-\frac{\tanh(\a\g h/2)}{\a\g h/2}\Big)\\ &\quad\qquad \ge \,\frac12\big(\frac{\t}{h^*}-\frac{\t}{h}\big)+\frac{\l}2m_\b^2\int\! \frac{\m (\rmd\a)}{\a}\Big(1-\frac{\tanh(\a\g h^*/2)}{\a\g h^*/2}\Big)\;,
\end{split}
\]
which is true for $C$ large enough (here $C$ is the constant appearing in the condition $h\ge C\g^{-2/3}$), simply because the expression in the first line is larger than $(C')^2\g^{2/3}$, with $C'$ proportional to $C$, while the right-hand side is smaller than $\frac12e(h^*)\le c_0\g^{2/3}$, with $c_0$ a universal constant, given by Eq.(\ref{00.6}). In conclusion, if $h\ge C\g^{-2/3}$, with $C$ large enough, 
\[
\widetilde e_h[\s]\ge e(h^*)+\frac1h\int_0^h\!\rmd x\, \widetilde F(\s(x)) +\frac12\big[e(h)-e(h^*)\big]\;,
\]
as desired. We are left with the case that $h\le C\g^{-2/3}$. Using the fact that $(\cdot,\cdot)_{\widetilde v_h}$ is a scalar product, we get,
\be
\label{5.33} 
\widetilde e_h[\s]\ge \frac1h\int_0^h\!\rmd x\, \widetilde F(\s(x)) +e(h)-\frac1h\|m_\b\|_{\widetilde v_h}\, \|\s-m_\b\|_{\widetilde v_h}\;.
\ee
Now a computation shows that  
\[
\begin{split}
& \frac1h\|m_\b\|_{\tilde v_h}^2\le C' (\g h)^{2}\le C'' \g^{2/3}\;, \\ & \|\s-m_\b\|_{\tilde v_h}^2\le C' (\g h)^{2} \|\s-m_\b\|_2^2\le C''\g^{2/3}\|\s-m_\b\|_2^2\;,
\end{split}
\]
where in the last line $\|\cdot\|_2$ is the standard $L_2$ norm on $[0,h]$. Plugging this into Eq.(\ref{5.33}) and using the fact that $\int_0^h\widetilde F(\s)=\frac{F(0)}{2m_\b^2}\|\s-m_\b\|^2_2$, we get,
\[
\widetilde e_h[\s]  \ge  \frac1{2h}\int_0^h\!\rmd x\, \widetilde F(\s(x)) +e(h) + \frac{F(0)}{4m_\b^2}h^{-1}\|\s-m_\b\|^2_2-C''\g^{2/3}h^{-1/2}\|\s-m_\b\|_2\;.
\]
The expression in the second line can be bounded from below by $\frac{(C'')^2m_\b^2}{F(0)}\g^{4/3}$, so that 
\[
\widetilde e_h[\s]\ge e(h^*)+\frac1{2h}\int_0^h\!\rmd x\, \widetilde F(\s(x)) +\big[e(h)-e(h^*)\big]-\frac{(C'')^2m_\b^2}{F(0)}\g^{4/3}\;,
\]
as desired. This concludes the proof of Eq.(\ref{c000}). 
\qed

%%%%%%%%%%%%%%%%%%%%%%%%%%%%%%%%%%%%%%%%%%%%%%%%%%%
\section{Conclusions}
\label{sec:conc}
%%%%%%%%%%%%%%%%%%%%%%%%%%%%%%%%%%%%%%%%%%%%%%%%%%%

By a combined use of coarse graining methods and reflection positivity, we studied the (approximate) minimizers of a one-dimensional non local free-energy functional with competing interactions, including short-range ferromagnetic interactions and long-range antiferromagnetic interactions. The short range potential is non-local of range 1 (and compact support), while the long range is a positive superposition of exponentials of range $\sim \g^{-1}\gg 1$. The competition among the two effects induces the quasi-minimizers to form a froth or foam: more precisely, they oscillate almost periodically, by alternating intervals where the profile is essentially equal to $+m_\b$ to intervals where it is essentially equal to $-m_\b$; the length of such intervals is all almost the same and of the order $\sim \g^{-2/3}$, which is intermediate between $1$ and $\g^{-1}$. The result is obtained by deriving estimates of the functional of interest  from above and below in terms of a ``sharp interface" functional with long range interaction, which can be studied by exact methods (reflection positivity). The bounds are extensive, i.e., proportional to the size of the interval on which the functional is defined, and subdominant in $\g$ with respect to the scale of the specific ground state energy. In this respect, our result is morally a sort of ``$\G$-convergence in infinite volume". 

It remains to be seen whether the infinite volume minimizers of our functional are exactly periodic or not. Consider for simplicity the case where both the long range potential and the short range one are pure exponentials, the first of range $\g^{-1}$ and the second of range 1. In this case, it may be possible to apply the methods of \cite{GLL3} to repeatedly reflect around the critical points of the potential $W(x)$ generated by the long range interaction. Why then couldn't we apply the methods of \cite{GLL3} to conclude that the minimizers are periodic? This may in principle be possible, but it would not be an immediate consequence of \cite{GLL3}, for a number of reasons:
\\
(1) If $|\phi|$ is not constant, as in our case, then $W(x)$ is not necessarily convex between two points 
where $\phi$ changes sign, contrary to what we had in \cite{GLL3}; therefore, it is not obvious that $W'(x)=0$ in a single point for each interval where $\phi$ has constant sign. 
\\
(2) Moreover, the proof of \cite{GLL3} is based on an explicit computation of the effective functional $f(p,q)$ obtained after reflections and on the remark that this function is jointly convex in $(p,q)$ (here $(p,q) = \{(p_i, q_i)\}$ are the lengths of the positive and negative parts of the function $\phi$ in each interval $I_i$ of the partition of $[0,L]$ induced by  the  critical points of $W$). If $|\phi|$ is not constant, as in our case, then it is not obvious to analyze the resulting effective functional, even if we could reflect around the zeros of $W'(x)$. 
\\
(3) Last but not least, the case that $J(x-y)$ is a pure exponential (or, more in general, it is reflection positive), on the one hand has the complication that $J$ is not of compact support, which complicates things a lot (one should generalize the analysis in \cite{Presutti}, which is used extensively in this paper, to  such a non trivial case). On the other hand, the assumption that $J$ is reflection positive is very restrictive and, as explained in the introduction, it has an independent interest to develop methods that combine perturbative (``cluster expansion"-like) methods with non-perturbative (reflection positivity) methods. 

\appendix

%%%%%%%%%%%%%%%%%%%%%%%%%%%%%%%%%%%
\section{On the convexity of $\widetilde F(t)$} 
\label{sec:A}
%%%%%%%%%%%%%%%%%%%%%%%%%%%%%%%%%%%

In this appendix we prove that $\widetilde F(t)=\frac{F(0)}{2m_\b^2}(|t|-m_\b)^2$ is such that $\widetilde F(t)\le \frac12 F(t)$. To this end, we recall the very definition of $F(t)$, which is
\[
\begin{split}
F(t) & = -\frac{\widehat J_0t^2}2+\frac1\b \Big(\frac{1+t}2\log\frac{1+t}2 + \frac{1-t}2\log\frac{1-t}2 \Big)\nonumber\\ & \quad + \frac{\widehat J_0m_\b^2}2-\frac1\b \Big(\frac{1+m_\b}2\log\frac{1+m_\b}2 + \frac{1-m_\b}2\log\frac{1-m_\b}2 \Big)\;,
\end{split}
\]
where $m_\b$ is the positive solution to $m_\b=\tanh(\b\widehat J_0 m_\b)$. The key estimate to be verified is $F''(m_\b)>2F(0)/m_\b^2$, that is,
\[
\begin{split}
-\widehat J_0 & +\frac1{2\b}\Big(\frac1{1+m_\b}+\frac1{1-m_\b}\Big) \\ & > \frac{2}{m_\b^2} \Big[\frac1\b\log\frac12+\frac{\widehat J_0}{2}m_\b^2-\frac1\b\Big(\frac{1+m_\b}{2}\log\frac{1+m_\b}{2} +\frac{1-m_\b}{2}\log\frac{1-m_\b}{2}\Big)\Big]\;.
\end{split}
\]
Multiplying both sides by $\b m_\b/2$ gives,
\[
\b m_\b\widehat J_0+\frac1{m_\b}\log\frac12<\frac{m_\b}{2(1-m_\b^2)}+\frac{1+m_\b}{2m_\b}\log\frac{1+m_\b}{2} +\frac{1-m_\b}{2m_\b}\log\frac{1-m_\b}{2}\;.
\]
Using the fact that $\b m_\b\widehat J_0=\frac12\log\frac{1+m_\b}{1-m_\b}$, we can rewrite the latter inequality as
\[ 
\frac{m_\b}{2(1-m_\b^2)}+\frac1{2m_\b}\log(1-m_\b^2)>0 \quad \Leftrightarrow\quad f(m_\b^2):=m_\b^2+(1-m_\b^2)\log(1-m_\b^2)>0\;,
\]
which is obviously verified, simply because $f(0)=0$ and $f'(x)=-\log(1-x)>0$ for any $x\in(0,1)$. Let us now turn to the proof that $F(x)\ge \frac{F(0)}{m_\b^2}(x-m_\b)^2$ for any $x\in[0,1]$. If $x\ge m_\b$ the claim follows immediately from the fact that $F''(x)=-\widehat J_0+\frac1\b\frac1{1-x^2}$ is an increasing function of $x$ in $[0,1]$. In particular, $F''(x)\ge F''(m_\b)$ for any $x\ge m_\b$, whence $F(x)\ge \frac12F''(m_\b)(x-m_\b)^2\ge \frac{F(0)}{m_\b^2}(x-m_\b)^2$ for any $x\in[m_\b,1]$. Regarding the interval $[0,m_\b]$, note that the function $g(x):=F(x)-\frac{F(0)}{m_\b^2}(x-m_\b)^2$ on $[0,m_\b]$ has two zeros at the boundaries, at $x=0$ and $x=m_\b$. Moreover, $g(x)$ is positive in $[0,\e]\cup[m_\b-\e,m_\b]$ for $\e$ small enough, simply because $g'(0)>0$, $g'(m_\b)=0$ and $g''(m_\b)>0$. Furthermore, $g''(x)$ is negative in zero, positive in $m_\b$ and monotonically increasing in $(0,m_\b)$: this also implies that $g'(x)$ goes monotonically from $g'(0)$ to its minimum (which is necessarily negative) and then increases monotonically from the minimum to $g'(m_\b)=0$; in particular, $g'(x)$ has a single zero in $(0,m_\b)$, which corresponds to the unique critical point of $g(x)$ in $[0,m_\b]$, which is a local maximum. In conclusion, $g(x)\ge 0$ in $[0,m_\b]$, which concludes the proof of the desired inequality.

%%%%%%%%%%%%%%%%%%%%%%%%%%%%%%%%%%%%%%%%%%%%%%%%%%%
\section{Proof of Lemma \ref{lem:1}}
\label{sec:B}
%%%%%%%%%%%%%%%%%%%%%%%%%%%%%%%%%%%%%%%%%%%%%%%%%%%

In the proof of Lemma \ref{lem:1} we will need the following sub-lemma.
\begin{lemma}
\label{lem:A1}
Under the assumptions of Lemma \ref{lem:1}, there exists a function $\c(x):B_i\to\{\pm m_\b\}$ with the following properties:
\\ \noindent
1) $\c(x)$ is measurable with respect to the partition $\{b_j\}$, i.e., $\c$ is constant on the interior of each small block $b_j$;
\\ \noindent
2) $\c(x)$ has a finite (i.e., independent of $\g$) number of jumps;
\\ \noindent
3) the $L_2$ distance between $\phi$ and $\c$ is finite, i.e.,
\be 
\label{A.1}
\int_{B_i}\!\rmd x\, \big[\phi(x)-\c(x)\big]^2\le \const \;,
\ee
for a suitable constant independent of $\g$.
\end{lemma}
Let us first show how to prove Lemma \ref{lem:1} assuming the validity of Lemma \ref{lem:A1}. The proof of Lemma \ref{lem:A1} will be then described below.  So, let us assume the validity of \eqref{A.1}. Given a small block $b_j$, we shall say that $b_j$ is {\it good} if $\big|\media{\phi-\c}_{b_j}\big|\le \g^\r$ and {\it bad} otherwise. The total number $N_1$ of bad blocks at a distance larger than $\ell_+/4$ from the boundary of $B_i$ can be bounded using that
\be 
\label{A.2}
\const \ge \int_{B_i}\! \rmd x\, \big[\phi(x)-\c(x)\big]^2\ge \ell_-\sum_{\substack{b_j\ {\rm bad} \\ \dist(b_j,\dpr B_i)\ge \ell_+/4}}\media{\phi-\c}_{b_j}^2\ge \ell_- N_1
\g^{2\r}\;,
\ee
that is $N_1\le \const  \g^{-2\r}$. Given a pair of contiguous blocks $(b_j,b_{j+1})$, we shall say that the pair is {\it good} if: (i) both its blocks are good; (ii) the value of $\c$ on both its blocks is the same. Note that the total number $N_2$ of bad pairs at a distance larger than $\ell_+/4$ from the boundary of $B_i$ can be bounded by $N_1+\const$, simply because $\c$ has a finite number of jumps. Using the bound on $N_1$ derived above, we have that $N_2\le \const  \g^{-2\r}$ as well. Now, by the pigeonhole principle, there will be at least one sequence of contiguous good pairs at a distance larger than $\ell_+/4$ from the boundary of $B_i$ with a number of blocks larger than $C \g^{-\d+2\r}$, for a suitable constant $C$ (here we used that the total number of contiguous pairs at a distance larger than $\ell_+/4$ from the boundary of $B_i$ is $\ell_+/(2\ell_-)=\const \cdot\g^{-\d}$). This completes the proof of Lemma \ref{lem:1}, modulo the proof of Lemma \ref{lem:A1}, which is described below.
\\ 
\noindent{\it  Proof of Lemma \ref{lem:A1}.} Let us first remark that there exists $c >0$ such that $F(t)\ge c \big(|t|-m_\b\big)^2$. If we fix $\k=m_\b/4$, we find that the set 
\be 
\label{A.3}
X_\k=\{x\in B_i\colon \big||\phi(x)|-m_\b\big|\ge \k\}
\ee
has finite measure, smaller than a constant independent of $\ell_+$. In fact,
\be 
\label{A.4}
|X_\k|\le \int_{X_\k}\!\rmd x\, \frac{\big(|\phi(x)|-m_\b\big)^2}{\k^2}\le \frac1{c\k^2}\int_{B_i}\!\rmd x\, F(\phi(x))\le \frac1{c\k^2} \mc E^{(0)}_{B_i}[\phi]\le \frac{2\t}{c\k^2}\;.
\ee
As a consequence, the number $N_3$ of blocks $b_j$ such that $|b_j\cap X_\k|\ge \ell_-/4$ is smaller than a suitable constant and 
\be 
\label{A.4a}
\int_{X_\k}\!\rmd x\, \big[\phi(x)-\c(x)\big]^2\le \const\;,\quad\sum_{b_j\colon  |b_j\cap X_\k|\ge \ell_-/4} \int_{b_j}\! \rmd x\, \big[\phi(x)-\c(x)\big]^2\le \const\;,
\ee 
independently of the choice of $\c$. Now, given a small block $b_j$, we set
\[
\begin{split}
b_j^\pm & =\{x\in b_j\colon \big|\phi\mp m_\b\big|<\k\}\;, \\ b_j' & = \begin{cases} b_j^- & {\rm if}\  |b_j^-|\le |b_j^+| \;, \\ b_j^+ & {\rm if}\  |b_j^+|< |b_j^-|\;, \end{cases} \qquad
b_j''=\begin{cases}b_j^+ & {\rm if}\  |b_j^-|\le |b_j^+| \;, \\ b_j^- & {\rm if}\ |b_j^+|< |b_j^-|\;.\end{cases}
\end{split}
\]
Calling $\underline b'$ the set of all the $b_j'$ such that $|b_j\cap X_\k|< \ell_-/4$, and noting that $J_1:=\min_{|x|\le\ell_-}J(x) =J(\ell_-)>0$ as $\ell_-<1$, we have,
\be 
\label{A.6}
\sum_{b_j'\in \underline b'}\int_{b_j'}\!\rmd x\, \big[\phi(x)-\c(x)\big]^2\le \frac{8}{3J_1\ell_-} \sum_{b_j'\in \underline b'}\int_{b_j'}\! \rmd x \int_{b_j''}\!\rmd y\, J(x-y) \big[\phi(x)-\phi(y)\big]^2\;,
\ee
independently of the choice of $\c$. In the previous bound we used that, if $b_j'\in \underline b'$, then $|b_j''|\ge 3\ell_-/8$. The right-hand side of \eqref{A.6} is finite, independently of $\g$, simply because it is smaller than $\mc E^{(0)}_{B_i}[\phi]$. Therefore, we are  left with 
\be 
\label{A.7}
\sum_{b_j''\in \underline b''}\int_{b_j''}\!\rmd x\, \big[\phi(x)-\c(x)\big]^2\;,
\ee
where $\underline b''$ is the set of all the $b_j''$ such that $|b_j\cap X_\k|<\ell_-/4$. If we choose $\c$ on $b_j$ of the same sign of $\phi$ on $b_j''$, we have that \eqref{A.7} is equal to 
\be 
\label{A.8}
\sum_{b_j''\in \underline b''}\int_{b_j''}\rmd x\, \big(|\phi(x)|-m_\b\big)^2\;,
\ee
which is smaller than $\const \mc E^{(0)}_{B_i}[\phi]$, as desired. We are left with proving that such a $\c$ has a finite number of jumps. In order to show this, let us note that the number of jumps in $\c$ is equal to the number $N_4$ of pairs of contiguous blocks $(b_j,b_{j+1})$ such that the sign $\o_j$ of $\phi$ on $b_j''$ is different from the sign $\o_{j+1}$ of $\phi$ on $b_{j+1}''$. Clearly $N_4$ is bounded above by the number of pairs containing a block such that $|b_j\cap X_\k|\ge \ell_-/4$ (which is finite, as proved above) plus a constant times
\[
\sum_{\substack{j\colon |b_j\cap X_\k|< \ell_-/4 \\ |b_{j+1}\cap X_\k|< \ell_-/4\;, \;\; \o_j\neq\o_{j+1}}} \int_{b_j''}\!\rmd x \int_{b_{j+1}''}\!\rmd y\, J(x-y)\big[\phi(x)-\phi(y)
\big]^2\;,
\]
which is smaller than $\mc E^{(0)}_{B_i}[\phi]$. 
\qed

%%%%%%%%%%%%%%%%%%%%%%%%%%%%%%%%%%%%%%%%%%%%%%%%%%%%
\section{Proof of Propositions \ref{prop:1} and \ref{prop:2}}
\label{sec:C}
%%%%%%%%%%%%%%%%%%%%%%%%%%%%%%%%%%%%%%%%%%%%%%%%%%%%

\subsection{Proof of Proposition \ref{prop:1}}
A preliminary result to prove Proposition \ref{prop:1} is Theorem \ref{thm:A1} below, which states in our context (part of) the results contained in \cite[Theorem 6.3.3.1]{Presutti}. Given $\phi\in L^\infty(\mathbb{R};[-1,1])$ and an interval $\mc I$, we denote by $\phi_\mc I$, $\phi_{\mc I^c}$ the restrictions of $\phi$ to $\mc I$ and $\mc I^c$, and define
\be
\label{2.6p}
\mc E_{\mc I}^{(0)}[\phi_\mc I|\phi_{\mc I^c}]= \mc E_{\mc I}^{(0)}[\phi_\mc I] +\frac12\int_{\mc I}\!\rmd x\int_{\mc I^c}\!\rmd y\, J(x-y)\big[\phi_\mc I(x)-\phi_{\mc I^c}(y)\big]^2\;.
\ee
We remark that, since we are assuming $J(x)=0$ for $|x|>1$, the energy $\mc E_{\mc I}^{(0)}[\phi_\mc I|\phi_{\mc I^c}]$ depends only on $\phi_{\tilde{\mc I}}$ where $\tilde{\mc I} =\{x\in\mathbb{R}\colon \dist (x,\mc I)\le 1\}$.
\begin{theorem}
\label{thm:A1} 
There are $\zeta_0, \kappa_0, \alpha$, and $c_\a$ all positive, so that for any $\zeta_1<\zeta_0$, $\ell_-<\kappa_0\zeta_1$, $\o\in \{\pm1\}$, and any bounded interval $\mc I$, measurable with respect to the partition $\{b_j\}$, the following holds.
\\  \noindent
1) If $|\media{\phi_{\mc I^c}}_{b_j}-\o m_\b|<\zeta_1$ for any $b_j\subset \tilde{\mc I}\setminus \mc I$ there is unique function $\psi_\mc I$ which minimizes $\mc E_{\mc I}^{(0)}[\phi_\mc I|\phi_{\mc I^c}]$ on the set of ``local equilibrium profiles'', i.e.\ those $\phi_\mc I$ with $|\media{\phi_\mc I}_{b_j}-\o m_\b|<\zeta_1$ for any $b_j\subset \mc I$.
\\ \noindent
2) $\psi_\mc I\in C^\infty(\mc I;[-1,1])$ and $|\psi_\mc I(x)-\omega m_\b| \le c_0 \rme^{-\alpha\,\mathrm{dist}\,(x,\mc I^c)}$.
\end{theorem}
\\ \noindent 
We can now proceed with the proof of Proposition \ref{prop:1}. Recalling the definition \eqref{2.6p}, we decompose
\be
\mc E_B^{(0)}[\phi] = \mc E_{B_1}^{(0)}[\phi_{B_1}] + \mc E_{\mc I_-}^{(0)}[\phi_{\mc I_-}|\phi_{\mc I_-^c}] + \mc E_{\mc I_+}^{(0)}[\phi_{\mc I_+}|\phi_{\mc I_+^c}]\;,
\label{3.7p}
\ee
where $\mc I_-$ is the union of the blocks $b_j$ such that $b_j\cap [a+1,a+2\log^2\ell-2] \ne\emptyset$, $\mc I_+$ is the union of the blocks $b_j$ such that $b_j\cap [a+\ell-2\log^2\ell+2,a+\ell-1] \ne \emptyset$, and $B_1 = B\setminus(\mc I_-\cup\mc I_+)$. We next define
$$
\hat\phi(x) = \begin{cases} \phi(x) & {\rm if} \quad x\in B_1\;,\\
\psi_{\mc I_\pm}(x) &{\rm if}\quad x\in \mc I_\pm\;,\end{cases}
$$
with $\psi_{\mc I_\pm}$ is the minimizer of local equilibrium profiles given in Theorem \ref{thm:A1}. By the decomposition \eqref{3.7p} we have,
\be 
\label{pb4}
\mc E_B^{(0)}[\hat \phi] \le \mc E_B^{(0)}[\phi]\;,
\ee
and, by Theorem \ref{thm:A1},
\begin{equation}
\label{bp6}
\begin{split}
&|\hat\phi(x) - \omega_-m_\b| \le c_\alpha \rme^{-\alpha\,(\log^2\ell-3)} \quad\forall\,x\in [a+\log^2\ell,a+\log^2\ell+1]\;, \\ & |\hat\phi(x) - \omega_+m_\b| \le c_\alpha \rme^{-\alpha\,(\log^2\ell-3)} \quad\forall\,x\in [a+\ell-\log^2\ell-1,a+\ell-\log^2\ell]
\end{split}
\end{equation}
(notice that the above intervals are at a distance not smaller that $\log^2\ell-3$ from the boundary of $\mc I_\pm$). The required function $\tilde\phi$ is now defined by
$$
\tilde\phi(x) = \begin{cases} \omega_-m_\b & {\rm if} \quad x\in (-\infty,a+\log^2\ell]\;, \\ \hat\phi(x) & {\rm if} \quad x\in (a+\log^2\ell,a+\ell-\log^2\ell)\;, \\ \omega_+m_\b & {\rm if} \quad x\in [a+\ell-\log^2\ell,\infty)\;.\end{cases}
$$
In fact, setting $\mc I=(a+\log^2\ell,a+\ell-\log^2\ell)$, by \eqref{bp6}, for a suitable $C_0>0$, 
\[
\begin{split}
\mc E_B^{(0)}[\hat \phi]  &  \ge  \mc E_{\mc I}^{(0)}[\hat \phi]  = \mc E_{\mc I}^{(0)}[\tilde \phi] \\ & \ge \mc E_{\mc I}^{(0)}[\tilde \phi] + \frac12\int_{\mc I} \!\rmd x \int_{B\setminus\mc I}\! \rmd y\, J(x-y)\big[\tilde\phi(x)-\tilde\phi(y)\big]^2 - C_0\, \rme^{-2\alpha\,\log^2\ell} \\  & = \mc E_B^{(0)}[\tilde \phi]-C_0\, \rme^{-2\alpha\,\log^2\ell}\;,
\end{split}
\]
where in the last equality we used that $\mc E_{B\setminus\mc I}^{(0)}[\tilde \phi]=0$. Therefore, by \eqref{pb4} the inequality \eqref{bp1} follows. Moreover, since $\tilde\phi(x)=\phi(x)$ for $x\in B_1\cap (a+2\log^2\ell-2, a+\ell-2\log^2\ell+2)$, $|\media{\tilde\phi_{\mc I_\pm}}_{b_j} - \media{\phi_{\mc I_\pm}}_{b_j}|\le 2\zeta_1$, and $|\mc I_\pm|\le  2\log^2\ell$, the bound \eqref{bp2} is straightforward. 
\qed 

\subsection{Proof of Proposition \ref{prop:2}}
\subsubsection{Proof of item (I)}
The only case in which the proof of Proposition \ref{prop:2} is really trivial is when $m\ge m_\b$ (or, similarly, $m\le -m_\b$). Indeed, both in the cases of open and periodic boundary conditions we can proceed as follows. To fix notation let us consider the case of open boundary conditions. Let $\phi_\l=m_\b+ \l[\phi-m_\b]_+$, where $\l\in[0,1]$ is chosen in such a way that $\media{\phi_\l}_B=m$. By construction, if $\phi\ge m_\b$ then $\phi\ge \phi_\l$, so that $\int_B\!\rmd x\, F(\phi)\ge \int_B\!\rmd x\,  F(\phi_\l)$. Now, by convexity, $\int_B\!\rmd x\,  F(\phi_\l)\ge |B|F(m)$, which proves item (I). The same argument proves item (III) under the assumption that $|m|\ge m_\b$.

\subsubsection{Proof of item (II)} Let us now turn to the proof of item (II). Without loss of generality, we consider the case $m_\b-\ell^{-1}\log^3\ell \le m < m_\b$ with $\phi_+$ boundary conditions. We consider the usual partition $\{b_j\}$ and we let $\phi$ be a configuration with energy $\mc E_B^{(0)}[\phi]\le \ell F(m)$. The key estimate to be proved is 
\be 
\label{c5}
\int_B\!\rmd x\, (\phi(x)-m_\b)^2 \le \const\frac{\log^6\ell}{\ell}\;.
\ee
Once this is known, then we immediately find that, for all $b_j\subset B$,
\[
|\media{\phi}_{b_j}-m_\b|\le \int_{b_j}\! \frac{\rmd x}{\ell_-} \, |\phi(x)-m_\b|\le \Big[\int_{b_j}\! \frac{\rmd x}{\ell_-} \, (\phi(x)-m_\b)^2 \Big]^{1/2} \le \const\frac{\log^3\ell}{\ell^{1/2}}\;.
\]
In other words, $\media{\phi}_{b_j}=m_\b+O(\ell^{-1/2}\log^3\ell)$ for all $b_j\subset B$. We now decrease the internal energy of $\phi$ in the following way. We start by replacing the configuration $\phi_{b_1}$ in $b_1$ (which is e.g. the first interval in $\{b_j\}$ in the ordering from left to right) by the minimizer of  $\mc E_{b_1}^{(0)}[\psi_{b_1} |\phi_{b_1^c}]$ under the constraint $\media{\psi_{b_1}}_{b_1}=\media{\phi}_{b_1}=:m_1$. By \cite[Theorem 6.4.1.1]{Presutti},  this minimizer is a smooth function $u_{b_1}(x)$ such that $|u_{b_1}(x)-m_1| \le \const\ell_-$ for any $x\in b_1$, provided $4\beta\|J\|_\infty\ell_-<1$. This amounts to replacing $\phi$ by the function $\psi^{(1)}(x)$ such that $\psi^{(1)}\big|_{b_1}=u_{b_1}$ and $\psi^{(1)}\big|_{b_1^c}=\phi\big|_{b_1^c}$. Then we repeat the procedure: we replace $\psi^{(1)}$ in $b_2$ by the minimizer of  $\mc E_{b_2}^{(0)}[\psi_{b_2} |\psi^{(1)}_{b_2^c}]$ under the constraint $\media{\psi_{b_2}}_{b_2}=\media{\phi}_{b_2}$, and so on. After a finite number of steps we end up with a new function $u(x)$ on $B$ such that $\mc E^{(0)}_B[u]\le \mc E^{(0)}_B[\phi]$, $\media{u}_{b_j}=\media{\phi}_{b_j}$ and $|u(x)-m_\b|\le \const\ell_-$. Therefore, if $\ell_-$ is small enough, $u(x)$ belongs to an interval on which $F(u)$ is convex, so that $\mc E^{(0)}_B[u]\ge \ell F(m)$, as desired. 

We are left with proving Eq.(\ref{c5}), under the assumption that $\mc E^{(0)}_B[\phi]\le \ell F(m) \le \const\ell^{-1}\log^6\ell$. We proceed as in Appendix \ref{sec:B}. Let $X:=\{x\colon \big||\phi(x)|-m_\b\big|\ge \z_0\}$, with $\z_0$ a small $O(1)$ constant. Obviously,
\[ 
\begin{split}
|X| \le \int_X\!\rmd x\, \frac{(|\phi(x)|-m_\b)^2}{\z_0^2} & \le \frac{m_\b^2}{\z_0^2 F(0)} \int_B\!\rmd x\,  F(\phi(x)) \\ & \le \frac{m_\b^2}{\z_0^2 F(0)}\mc E^{(0)}_B[\phi] \le \const\frac{\log^6\ell}{\ell}\;.
\end{split}
\]
Given $b_j$, let $b_j^\pm =\{x\in b_j\colon \big|\phi\mp m_\b\big|<\z_0\}$ and
\[
b_j' = \begin{cases} b_j^- & {\rm if}\  |b_j^-|\le |b_j^+| \;, \\ b_j^+ & {\rm if}\  |b_j^+|< |b_j^-|\;, \end{cases} \qquad b_j''=\begin{cases}b_j^+ & {\rm if}\  |b_j^-|\le |b_j^+| \;, \\ b_j^- & {\rm if}\ |b_j^+|< |b_j^-|\;.\end{cases}
\]
Now
\bea
\label{c7}
&& \int_B\!\rmd x\, \big(\phi(x)-m_\b\big)^2 \nonumber \\ &&=\int_X\!\rmd x\,\big(\phi(x)-m_\b\big)^2 + \sum_j\Big[\int_{b_j'}\!\rmd x\, \big(\phi(x)-m_\b\big)^2+\int_{b_j''}\!\rmd x\, \big(\phi(x)-m_\b\big)^2\Big] \nonumber \\ && \le 2|X|+\sum_j\Big[C\int_{b_j'}\!\rmd x \int_{b_j''}\!\rmd y\, \big(\phi(x)-\phi(y)\big)^2 J(x-y)+\int_{b_j''}\!\rmd x\, \big(\phi(x)-m_\b\big)^2\Big] \nonumber\\ && \le C'\frac{\log^6\ell}{\ell}+\sum_j\int_{b_j''}\!\rmd x\, \big(\phi(x)-m_\b\big)^2 \;,
\eea
where in the first inequality we used the fact that
\[ 
\begin{split}
\int_{b_j'}\!\rmd x\, \big(\phi(x)-m_\b\big)^2 & \le  |b_j'|(2m_\b+\z_0)^2 \\ & \le \frac{(2m_\b+\z_0)^2}{(2m_\b-2\z_0)^2}\int_{b_j'}\!\rmd x \int_{b_j''}\,\frac{\rmd y}{|b_j''|}\, \big(\phi(x)-\phi(y)\big)^2\frac{J(x-y)}{J(\ell_-)}\;,
\end{split}
\]
with $|b_j''|\ge \frac12(|b_j''|+|b_j'|) \ge \frac12(\ell_--\const\frac{\log^6\ell}{\ell})$. Let now $\o_j=\sign\,\phi(x)\big|_{b_j''}$ and let $N$ be the number of jumps from $\o_j=+$ to $\o_{j+1}=-$ or viceversa. We have,
\[ 
N\le \const\sum_j\int_{b_j''}\!\rmd x \int_{b_{j+1}''}\!\rmd y\, J(x-y)\big(\phi(x)-\phi(y)\big)^2 \le \const\frac{\log^6\ell}{\ell}\;,
\]
which means that $N=0$ for $\ell$ large enough. In other words, $\o_j\equiv +$ for any $j$, and, therefore, $\sum_j\int_{b_j''}\!\rmd x\,\big(\phi(x)-m_\b\big)^2= \sum_j\int_{b_j''}\!\rmd x\,\big(|\phi(x)|-m_\b\big)^2\le \const\ell^{-1}\log^6\ell$. Plugging this into Eq.(\ref{c7}) we get the desired result.

\subsubsection{Proof of item (III)}
We are left with the analysis of the case $|m|\le m_\b$ with periodic boundary conditions: recall in fact that the case $|m|\ge m_\b$ with periodic boundary conditions was studied above, together with the proof of item (I). We adapt to the present context the strategy used in \cite{CCELM2} where the energy on a torus in dimension greater than one is considered. A key observation which simplifies the analysis in our one-dimensional case is that for a lower bound on the energy it is sufficient to consider symmetric monotone configurations. This follows by rearrangement inequalities on the torus, which are valid under the assumption that $J$ is monotone \cite{BT,CCELM1}. More precisely, to show this fact we rewrite the energy in the form,
\[
\mc E_{\mc T}^{(0);\mathrm{per}}[\phi]=\int_{\mc T}\! \rmd x\,  W(\phi(x))-\frac12\int_{\mc T}\! \rmd x\int_{\mc T}\! \rmd y\, J(x-y)\phi(x)\phi(y)\;,   
\]
with $W(t) = F(t) + \frac 12  \widehat J_0 t^2$. We identify $\mc T$ with the interval $[-\frac\ell2,\frac\ell2]$ with periodic boundary conditions and we recall that the function $\phi$ on $\mc T$ is said symmetric monotone decreasing if $\phi(x)=\phi(-x)$ and its restriction to $[0,\ell/2]$ is  monotone decreasing. The symmetric monotone decreasing rearrangement of a function $\phi$ is now defined as the symmetric monotone decreasing function $\phi_*$ on $\mc T$ such that $\{x\colon \phi_*(x) > t\}$ has the same measure of $\{x\colon \phi(x) > t\}$ for all $t > 0$. The function $\phi_*$ is uniquely defined, except for sets of measure zero \cite{BT}. By the definition of $\phi_*$ it follows immediately that $\int_{\mc T}\! \rmd x\,  W(\phi(x)) = \int_{\mc T}\! \rmd x\,  W(\phi_*(x))$. Moreover, since $J$ is monotone, by \cite[Theorem 2]{BT}, 
\[
\frac12\int_{\mc T}\! \rmd x\int_{\mc T}\! \rmd y\, J(x-y)\phi(x)\phi(y) \le \frac12\int_{\mc T}\! \rmd x\int_{\mc T}\! \rmd y\, J(x-y)\phi_*(x)\phi_*(y)\;.
\] 
Therefore, $\mc E_{\mc T}^\mathrm{per}[\phi] \ge \mc E_{\mc T}^\mathrm{per}[\phi_*]$, so that it is enough to consider only symmetric monotone decreasing profiles $\phi$. Moreover, since the energy $\mc E_{\mc T}^\mathrm{per}[\phi]$ is invariant under reflection $\phi\to (-\phi)$, it is enough to consider positive average, $m\in [0,m_\b]$. 

As $F(m)$ is monotone decreasing on $[0,m_\b]$ and $\ell F(m_\b - c\ell^{-1/2}) = \frac 12 F''(m_\b) c^2 + O(\ell^{-1/2})$, we have, for any $\ell$ large enough,
\be
\label{cc*}
\begin{split}
& \ell F(m)>\frac94\tau \quad \forall\, m\in  [0,m_\b-C_*\ell^{-1/2}]\;,\\
& \ell F(m) < \frac74\tau \quad\forall\, m\in [m_\b-c_*\ell^{-1/2}, m_\b]\;.
\end{split}
\ee
where 
\[ 
c_*=\sqrt{\frac{3\t}{F''(m_\b)}}\;, \qquad C_*=\sqrt{\frac{5\t}{F''(m_\b)}}\;.
\]
We first consider the case $m\in[0,m_\b-C_*\ell^{-1/2}]$. Since the instanton converges exponentially fast to $\pm m_\b$ as $x\to\pm\infty$, as remarked in the lines after Eq.(\ref{2.inst}), it is easy to check that the trial function $\phi_m(x) = q(|x|-z)$, $x\in [-\ell/2,\ell/2]$, with $z$ such that $\media{\phi_m}_{\mc T} = m$, has energy $\mc E_{\mc T}^{(0);\mathrm{per}}[\phi_m] = 2\tau + O(\rme^{-2\alpha\,\log^2\ell})$, with $\alpha$ bounded above by the rate of the exponential convergence of $q(x)$ to $\pm m_\b$. Without loss of generality, we shall assume hereafter $\alpha$ to be the same as in Theorem \ref{thm:A1}. Now let $\phi$ be an absolute minimizer of the energy. As remarked above, $\phi$ can be assumed to be a monotone symmetric profile with $\mc E_{\mc T}^{(0);\mathrm{per}}[\phi] \le 2\tau + O(\rme^{-2\alpha\,\log^2\ell})$. As usual, we consider the partition $\{b_j\}$ of $\mc T$ and we fix a small tolerance $\zeta_0$, of order 1 with respect to $\g$. We may argue as in the proof of Lemma \ref{lem:1} to conclude that the number $N$ of bad intervals where $\big||\media{\phi}_{b_j}|-m_\b\big|\ge\zeta_0$ is bounded by $\const\z_0^{-2}$; similarly,
\be
\label{c11}
|X_{\z_0}|\le \const\z_0^{-2}\;,\qquad {\rm where}\qquad X_{\z_0}:=\{x\colon \big||\phi(x)|-m_\b\big|\ge \z_0\}\;.
\ee
Let us denote by $\mathcal{I}_\pm$ the set where $|\media{\phi}_{b_j}\mp m_\b|<\zeta_0$. Since $\phi$ is monotone symmetric, each of the sets $\mathcal{I}_\pm$ is at most the union of two intervals on $\mc T$. Moreover, since $\media{\phi}_{\mc T}=m$, 
\be 
\label{c10bis}
\left||\mathcal{I}_\pm| - \frac{m_\b\pm m}{2m_\b}\ell\right| \le \const  \left(\zeta_0\ell+\frac{1}{\zeta_0^2}\right)\;.
\ee
As $m\ge 0$, the above bound implies $|\mathcal{I}_+|\ge \ell/4$ for $\zeta_0$ small enough. We remark that the same estimate is useless to get an upper bound on $|\mathcal{I}_-|$ because $(m_\b-m)\ell$ can be of order $C_* \ell^{1/2}$. On the other hand, also $|\mathcal{I}_-|$ cannot be too small, otherwise the profile $\phi$ would have an energy close to $\ell F(m)$, which is larger than $\frac94\tau$ by \eqref{cc*}. Indeed, suppose that $|\mathcal{I}_-|\le \ell^{1/4}$, definitively for any $\ell$ large enough. Then $|\mathcal{I}_+|=\ell-|\mc I_-|-|X_{\z_0}|\ge\ell-\ell^{1/4}-\const\z_0^{-2}\ge \ell (1-C \ell^{-3/4})$. Since $\phi$ is monotone symmetric, this implies that also the region $\mc T_+ = \{x\in\mc T\colon \phi(x)\ge m_\b-\zeta_0\}$ has measure larger than $\ell(1-C \ell^{-3/4})$. Therefore, choosing $\zeta_0$ small enough to have $F(t)$ convex for $t\ge m_\b-\zeta_0$,
\be
\label{c8}
\mc E^{(0);\mathrm{per}}_{\mc T}[\phi]\ge \int_{\mc I_+}\!\rmd x\, F(\phi(x))\ge |\mc I_+|F(m_+)\;,
\ee
where $m_+\ell (1-C\ell^{-3/4}) \le m_+|\mc I_+|=m\ell-\int_{\mc I_-\cup X_{\z_0}}\!\rmd x\, \phi(x)\le m\ell(1+C'\ell^{-3/4})$, which implies that $m_+\le m+C''\ell^{-3/4}\le m_\b-\ell^{-1/2}C_*(1-C'''\ell^{-1/4})<m_\b$. Plugging this into Eq.(\ref{c8}) gives,
\be
\label{c14}
\begin{split}
\mc E^{(0);\mathrm{per}}_{\mc T}[\phi]&\ge \ell(1-C\ell^{-3/4}) F\big(m_\b-\frac{C_*}{\ell^{1/2}}(1-C'''\ell^{-1/4})\big)\\ & \ge \frac{\ell }2F''(m_\b)\frac{C_*^2}{\ell} (1-\tilde C\ell^{-1/4})\ge \frac94\t\;,
\end{split}
\ee
which is in contradiction with the assumption that $\mc E^{(0);\mathrm{per}}_{\mc T}[\phi]\le 2\t +O(\rme^{-2\a\log^2\ell})$. Hence $|\mathcal{I}_-|> \ell^{1/4}$, definitively for any $\ell$ large enough. Let us pick two intervals $J_+$ and $J_-$, contained in $\mc I_+\cap [0,\ell/2]$ and in $\mc I_-\cap [0,\ell/2]$, respectively, both of length of order $\ell^{1/4}$ at least and at a distance $>\const\ell^{1/4}$ from the boundary of $\mc I_+\cap [0,\ell/2]$ and of $\mc I_-\cap [0,\ell/2]$, respectively. By Theorem \ref{thm:A1}, we can improve the internal energy $\mc E^{(0);\mathrm{per}}_{\mc T}[\phi]$ by replacing $\phi$ on $J_+$ by the minimum of $\mc E^{(0)}_{J_+}(\psi_{J_+}|\phi_{J_+^c})$, and similarly for $J_-$. We denote by $\hat \phi$ the function obtained from $\phi$ after the two replacements in $J_+$ and $J_-$. In all the points of $\tilde J_+=\{x\in J_+\colon {\rm dist}(x,J_+^c)\ge \log^2\ell\}=[a_+,b_+]$, we have that $|\hat \phi-m_\b|\le c_\a \rme^{-\a\log^2\ell}$. Therefore, if we further replace $\hat \phi$ on $[b_+,\ell/2]$ by $m_\b$, we further decrease the energy, up to a possible error of the order $\rme^{-\a\log^2\ell}$; we denote by $\tilde\phi$ the resulting modified function. Similarly, we can define $\tilde J_-=[a_-,b_-]=\{x\in J_-\colon {\rm dist}(x,J_-^c)\ge \log^2\ell\}$  and we can further decrease the energy (up to errors of the order $\rme^{-\a\log^2\ell}$) by changing $\tilde \phi$ to a new function $u(x)$ that is equal to $-m_\b$ on $[0,a_-]$. In conclusion, we replaced the original function $\phi$ by a new function $u$ that is constantly $-m_\b$ and $m_\b$ in long intervals of the order at least $\ell^{1/4}$ that are well separated among each other, by a distance of the order at least $\ell^{1/4}$. At this point, by making use of the results on the infinite volume problem with $(-,+)$ boundary conditions, see Eq.(\ref{2.inst}) and following lines, $\mc E_{\mc T}^{(0);\mathrm{per}}[\phi]\ge \mc E_{\mc T}^{(0);\mathrm{per}}[u]+O(\rme^{-2\a\log^2\ell}) \ge  2\tau+O(\rme^{-2\a\log^2\ell})$. Eq.\eqref{i4} for $m\in [0,m_\b-C_*\ell^{-1/2}]$ is thus proved.

\medskip
We now consider the more delicate case $m\in [m_\b-C_*\ell^{-1/2},m_\b]$. Given a monotone symmetric profile $\phi(x)$ with average $m\in [m_\b-C_*\ell^{-1/2},m_\b]$, we set
\begin{equation}
\label{pp1}
\varrho =m_\b - m\;, \qquad h_\pm=\pm m_\b\mp \varrho^{1/3}
\end{equation}
and, by slicing $\mc T$ at the $h_\pm$ levels of $\phi$, we define the sets,
\begin{equation}
\label{pp2}
\begin{split}
H_- & =\{x\in\mc T\colon \phi(x) \le h_-\}\;,\qquad  S=\{x\in\mc T\colon h_-\le \phi(x) \le h_+\}\;, \\  H_+ & =\{x\in\mc T\colon \phi(x) \ge h_+\}\;.
\end{split}
\end{equation}
Since $\phi$ is monotone symmetric $H_\pm$ are connected subsets of the torus. In particular, if $H_-$ is nonempty then it is an interval centered at $x=0$; we denote its size by $2R$, i.e., $H_-=[-R,R]$. The key technical result is the following.
\begin{lemma}
\label{lem:B1} There exists a positive constant $c$ such that, given a function 
$\phi$ on $\mc T$ with average $m=m_\b-\varrho$, $\varrho\in (0,C_*\ell^{-1/2}]$ and 
$\mc E_{\mc T}^\mathrm{per}[\phi] < \ell F(m)$, then $H_-$ is not empty and 
\begin{equation}
\label{H-ge}
R \ge c \varrho^{-1}\;.
\end{equation}
\end{lemma}
Once this lemma is proved, we are essentially done. In fact, either the minimum of $\mc E^{(0);\mathrm{per}}_{\mc T}[\phi]$ over $\phi$ is {\it equal} to $\ell F(m)$, or $\min_\phi\mc E^{(0);\mathrm{per}}_{\mc T}[\phi]<\ell F(m)$, in which case we can apply Lemma \ref{lem:B1} to conclude that $R\ge \const\ell^{1/2}$. This means in particular that the set $\mc I_-$ defined after Eq.(\ref{c11}) has measure larger than $\const\ell^{1/2}$. We already know from  Eq.(\ref{c10bis}) that $\mc I_+$ has measure larger than $\const\ell$. Then we can proceed as explained after Eq.(\ref{c14}) to conclude that, if $\mc E^{(0);\mathrm{per}}_{\mc T}[\phi]<\ell F(m)$, then $\mc E^{(0);\mathrm{per}}_{\mc T}[\phi]\ge 2\t +O(\rme^{-2\a\log^2\ell})$, which is of course possible only for certain values of $m$. This would conclude the proof of item (III). Of course,  we are left with proving Lemma \ref{lem:B1}.\\

\noindent{\it Proof of Lemma \ref{lem:B1}}. We start by deriving a couple of useful a priori upper bounds on the size of $|S|$ and $|H_-|$, namely
\begin{equation}
\label{Sle}
|S| \le c_1 \varrho^{4/3}\ell\;,
\end{equation}
\begin{equation}
\label{H-le}
R \le c_1 \varrho\ell\;.
\end{equation}
for a suitable $c_1>0$. Hereafter we assume that $\mc E^{(0);\mathrm{per}}_{\mc T}[\phi]\le\ell F(m)$. By the definition of $S$ and the explicit form of $F$ we have  $F(\phi(x))\ge F(h_+)$ for any $x\in S$, whence
\[
\ell F(m) \ge \mc E_{\mc T}^\mathrm{per}[\phi] \ge |S| F(h_+)\;.
\]
Moreover, again by the explicit form of $F$, there is $K>1$ such that
\begin{equation}
\label{beta}
\frac 1K (t-m_\b)^2 \le F(t) \le K (t-m_\b)^2 \qquad \forall\,t\in[0,m_\b]\;. 
\end{equation}
Therefore, by \eqref{pp1} the bound \eqref{Sle} follows, with $c_1=K^2$ for any $\varrho\in [0, m_\b]$. To prove \eqref{H-le} we start with the obvious identity,
\begin{equation}
\label{SHH}
\ell(m_\b-\varrho) = \int_{H_-}\! \rmd x\, \phi(x) + \int_S\! \rmd x\, \phi(x) +\int_{H_+}\! \rmd x\, \phi(x)\;,
\end{equation}
which implies
\[
\ell(m_\b-\varrho) \le h_- |H_-| + (|S|+|H_+|) = \ell h_- + (1-h_-) (|S|+|H_+|)\;,
\]
where in the upper bound we used the definition of $H_-$ and that $\phi\le 1$, while in the equality that $|H_-| = \ell - |S|-|H_+|$. We have already proved that $|S| \le K^2\varrho^{4/3}$, therefore, for $\varrho$ small enough,
\begin{equation}
\label{H+}
|H_+| \ge \frac{\ell(2m_\b-\varrho^{1/3}-\varrho)}{1+m_\b-\varrho^{1/3}} - K^2 \varrho^{4/3} \ge \frac{m_\b\ell}{1+m_\b}\;. 
\end{equation}
On the other hand, by the assumption on the energy of $\phi$ and using \eqref{beta},
\[
\begin{split}
\ell K\varrho^2 & \ge \ell F(m) \ge \mc E_{\mc T}^\mathrm{per}[\phi] \ge \int_{H_+}\! \rmd x\, F(\phi(x)) \\ & \ge \frac 1K \int_{H_+}\! \rmd x\, [\phi(x)-m_\b]^2 \ge \frac{|H_+|}K \left(\media{\phi}_{H_+} - m_\b\right)^2\;,
\end{split}
\]
whence, using also the lower bound \eqref{H+} on $|H_+|$,
\[
\left|\media{\phi}_{H_+} - m_\b\right| \le K \sqrt{\frac{1+m_\b}{m_\b}} \, \varrho=:K_1\varrho\;.
\]
Finally, plugging this estimate in \eqref{SHH},
\[
\begin{split}
\ell(m_\b-\varrho) & \le h_-|H_-| + h_+|S| + (\ell -|S|-|H_-|) \media{\phi}_{H_+} \\ & \le  h_-|H_-| + h_+|S| + (m_\b+K_1\varrho) (\ell -|S|-|H_-|)\;,
\end{split}
\]
that is, by the definition \eqref{pp1} of $h_\pm$, 
\[
(2m_\b-\varrho^{1/3}+K_1\varrho)\, |H_-| \le (1+K_1)\varrho\ell -|S|(\varrho^{1/3}+K_1\varrho) \le (1+K_1)\varrho\ell\;.
\]
The last bound implies $|H_-| \le (1+K_1)m_\b^{-1}\varrho\ell$ for any $\varrho$ small enough. The bounds \eqref{Sle} and \eqref{H-le} are thus proved with $c_1=\max\{(1+K_1)m_\b^{-1};K^2\}$ for any $\varrho\in (0,C_*\ell^{-1/2}]$ if $\ell$ is sufficiently large.\\

We are now ready for the proof of Eq.(\ref{H-ge}). For $\phi$ as in the hypothesis of the lemma we can rewrite the energy in the form,
\begin{equation}
\label{EG}
\mc E_{\mc T}^\mathrm{per}[\phi] = \ell F(m) + \mc G_{\mc T}[\phi]\:, 
\end{equation}
where
\begin{equation}
\label{GG}
\mc G_{\mc T}[\phi]=\int_{\mc T}\! \rmd x\, G(\phi(x))+\frac14\int_{\mc T}\! \rmd x\int_{\mc T}\! \rmd y\, J(x-y)\big[\phi(x)-\phi(y)\big]^2\;,
\end{equation}
with
\begin{equation}
\label{G}
G(t) = F(t) - F(m) - F'(m) (t-m), \qquad t\in [-1,1]\;.
\end{equation}
By the explicit form of $F(t)$ it is not difficult to prove that there is a  positive constant $K_2$ such that for any $\varrho$ small enough the following holds. The function $G(t)$ has a double well shape, with an absolute minimum at a point $t=m_*<0$ and a local minimum at $t=m$. Moreover $G(t)$ has exactly three zeros, precisely at $t=m$ and $t=m_\pm$, with $m_-<m_+<0$. Finally, 
\begin{equation}
\label{K2}
|m_*+m_\b| \le K_2\varrho\;, \qquad |m_\pm+m_\b| \le K_2\varrho^{1/2}\:, \qquad G(m_*) \ge -K_2\varrho\;.
\end{equation}
In particular, $G(t)<0$ if and only if $t\in (m_-,m_+)$ whence, by \eqref{EG}, \eqref{GG}, and the assumption $\mc E_{\mc T}^\mathrm{per}[\phi] < \ell F(m)$ we conclude that $D := \{x\in \mathcal{T}\colon m_-<\phi(x)<m_+\}$ is not empty. Moreover, by the first inequality in \eqref{K2} and the definition \eqref{pp2} of $H_-$, if $\varrho$ is small enough then $D\subset H_-$, hence also $H_-$ is not empty. As a further preliminary step towards the proof of \eqref{H-ge}, we show that $2R>\frac 12$ for any $\varrho$ small enough. Let $H = \{x\in \mathcal{T}\colon \phi(x)<m_+\}$. By definitions \eqref{pp2}, \eqref{GG}, and the bounds \eqref{K2} we have, 
\[
\begin{split}
\mc G_{\mc T}[\phi] & \ge \int_H\! \!\rmd x\, G(\phi(x))+\frac12\int_H\! \rmd x \int_{S\cup H_+} \! \rmd y\,  J(x-y)\big[\phi(x)-\phi(y)\big]^2 \\ & \ge -  |H| K_2\varrho +  \frac{(\varrho^{1/3}-K_2\varrho^{1/2})^2}2 \int_H\! \rmd x \int_{S\cup H_+} \! \rmd y\,  J(x-y) \\ & \ge  -  |H| K_2\varrho + K_3\frac{(\varrho^{1/3}-K_2\varrho^{1/2})^2}2 \int_H\! \rmd x \int_{S\cup H_+} \! \rmd y\,  \mathcal{X}_{|x-y|\le 1/2} \;,
\end{split}
\]
where  
\begin{equation}
\label{K3}
K_3 = \min_{|\xi|\le 1/2} J(\xi) = J(1/2) >0\;. 
\end{equation}
Now, if $|H_-|=2R \le \frac 12$, the intersection of the interval $\left[x-\frac 12, x+\frac 12\right]$ with $S\cup H_+$ is larger than $\frac 14$ for any $x\in H$, whence
\[
\mc G_{\mc T}[\phi] \ge  -  |H| K_2\varrho + |H| K_3\frac{(\varrho^{1/3}-K_2\varrho^{1/2})^2}8\;.
\]
Since the right-hand side is strictly positive for $\varrho$ small, we conclude that the assumption $\mc G_{\mc T}[\phi]<0$ implies $2R>\frac 12$ for any $\varrho$ small enough. 

To improve this lower bound on $R$ we observe that, by the last estimate in \eqref{K2} and as $D\subset H_-$, 
\[
\begin{split}
\mc G_{\mc T}[\phi] & \ge \frac14\int_{H_-}\! \rmd x\int_{H_+}\! \rmd y\,  J(x-y)\big[\phi(x)-\phi(y)\big]^2 -2RK_2\varrho \\ & \ge K_3(m_\b-\varrho^{1/3})^2 \int_{H_-}\! \rmd x \int_{H_+}\! \rmd y\,  \mathcal{X}_{|x-y|\le 1/2}  -2RK_2\varrho\;,
\end{split}
\]
where in last inequality we used the definitions \eqref{pp1}, \eqref{pp2}, and \eqref{K3}. We now notice that by \eqref{H+} with $\ell\ge 1$, as $|H_-|\ge \frac 12$ and $|S|\le K^2\varrho^{4/3}$, the double integral on the right-hand side is bounded from below by some positive constant for any $\varrho$ small enough. Therefore, for a suitable constant $K_4>0$, if $\varrho\in (0,C_*\ell^{-1/2}]$ and $\ell$ is sufficiently large, 
\[
\mc G_{\mc T}[\phi] \ge K_4 - 2RK_2\varrho\;.
\]
But $\mc G_{\mc T}[\phi]<0 $ so that \eqref{H-ge} follows with $c_2 = K_4/(2K_2)$.
\qed

\medskip
As observed above, right after its statement, once that the Lemma \ref{lem:B1} is proved, the proof of Proposition \ref{prop:2} is concluded as well. 
\\
{\bf Acknowledgements.} A.G. gratefully acknowledges financial support from the ERC Starting Grant CoMBoS-239694. A.G. would also like to thank J. Lebowitz and E. Lieb: this work is part of a joint project started together and several of the ideas used in this paper are due to their suggestions and to common discussions on the issue of periodic minimizers. Moreover, we thank J. Lebowitz for suggesting us this problem and for encouragement to complete it. Finally, we warmly thank E. Presutti and A. de Masi for several illuminating discussions and for some important technical suggestions about the definition of the coarse graining procedure.

\end{document}